\documentclass[reprint,
twocolumn,
amsmath,
amssymb,
aps,
physrev]{revtex4-2}
\usepackage{graphicx}
\usepackage{dcolumn}
\usepackage{bm}
\usepackage{hyperref} 
\hypersetup{ 
	colorlinks=true,
	linkcolor=blue,
	filecolor=magenta,      
	urlcolor=blue,
	citecolor=red
}
\usepackage[justification=justified]{caption}
\usepackage{labelschanged}
\usepackage{xcolor} 
\newcommand{\comment}[1]{}  

\newif\ifhighlight
\highlightfalse    

\newcommand{\hl}[1]{%
	\ifhighlight
	\textcolor{cyan}{#1}
	\else
	#1
	\fi
}

\begin{document}
	
	\preprint{APS/123-QED}
	
	\title{\textbf{Carrier-envelope phase and pulse shape effects on vacuum pair production in asymmetric electric fields with bell-shaped envelopes}}%
	
	\author{Abhinav Jangir}
	\email{2022rpy9087@mnit.ac.in}
	\author{Anees Ahmed}
	\email{anees.phy@mnit.ac.in}
	\affiliation{
		Department of Physics, Malaviya National Institute of Technology Jaipur, Jaipur, Rajasthan, India
	}
	
	\date{\today}
	
	\begin{abstract}
		We investigate the combined effects of carrier-envelope phase and laser pulse shape on electron-positron pair production in the presence of an external time-dependent asymmetric electric field by solving the quantum Vlasov equation. We analyze how the pulse asymmetry, the envelope type (Gaussian, Lorentzian and Sauter), and the carrier-envelope phase jointly influence the momentum distribution and the total number of produced pairs per unit volume. Our results show that pair production exhibits extreme sensitivity to both the degree of the temporal asymmetry and the steepness of the envelope on either side of the pulse. These effects are qualitatively explained through a turning-point analysis for the non-analytic electric field using a regularization scheme. We observe that multiphoton pair production dominates the Schwinger mechanism in the case of a long falling-pulse asymmetry. For a short falling pulse with a flat-topped profile, pair production is further facilitated. We demonstrate that the density of produced pairs can be enhanced by two to three orders of magnitude by choosing certain field parameters.
	\end{abstract}
	
	\maketitle
	\clearpage
	
	\section{\label{sec:Intro}Introduction} 
	Dirac's prediction of the positron \cite{dirac1928quantum} paved the way for Sauter to demonstrate that the vacuum can decay into electron-positron pairs in the presence of a very strong static electric field \cite{sauter1931behavior}. Later, in 1951, Schwinger derived the pair production rate in a constant electric field using the proper-time method \cite{schwinger1951gauge}, identifying the critical field strength $E_\text{cr} = {m^2c^3}/{e\hbar} \approx 1.3 \times 10^{18} \mathrm{V/m}$ as the threshold for producing observable pairs, where $m$ and $e$ are the electron mass and fundamental charge respectively. The field strength $E_\text{cr}$ corresponds to a laser intensity $\sim 10^{29}$ W/cm$^2$, far exceeding present laser capabilities. Although chirped-pulse amplification (CPA) \cite{strickland1985compression} technology has dramatically increased achievable laser intensities, researchers at CoReLS demonstrated a record-high peak laser intensity exceeding $10^{23}$ W/cm${^2}$ \cite{yoon2021realization} by tightly focusing a petawatt-class laser pulse. This is still far beyond the critical laser intensity. Consequently, direct experimental verification of the Schwinger effect is still pending. 
	The advent of the Extreme Light Infrastructure (ELI) \cite{ELI}, the X-ray free electron laser (XFEL) \cite{XFEL, ringwald2001pair} and eXwatt Center of Extreme Light Studies (XCELS) \cite{khazanov2023exawatt} may enable experimental investigations of QED vacuum decay into electron-positron pairs in the near future. 
	
	The non-perturbative non-equilibrium process of electron-positron pair production has been studied using a variety of approaches. These \hl{include numerical methods to directly solve the Dirac equation \cite{gitman1996, linder2015, wollert2015, aleksandrov2016, lv2018}, WKB approximation \cite{popov1972pair, dipiazza2021, oertel2019, kohlfurst2019wkb, dumlu2010stokes}, effective Lagrangian techniques \cite{dunne2005fields, dunneharris2023, gies2017}, worldline instanton methods \cite{dunne2005worldline, ilderton2015worldline}, the Wigner formalism \cite{olugh2019pair,chen2024asymmetric, hebenstreit2010DHWvsQKE, HebenstreitPhD, BlinnePhD, li2015effects}, and the quantum kinetic formalism (QKF) \cite{kluger1998quantum, bloch1999pair, alkofer2001pair, blaschke2006pair, dumlu2009equivalence, orthaber2011momentum, aleksandrov2020generalQKE, aleksandrov2024generalQKE}. Each of these approaches provides the pair production rate, while having its own advantages and disadvantages. QKF} has the advantage of yielding the momentum distribution of the produced pairs. Being fully non-perturbative, QKF naturally incorporates contributions from both multiphoton absorption (a perturbative effect \cite{mocken2010nonperturbative, ruf2009pair, kohlfurst2014effective}) and the Schwinger effect (a purely non-perturbative effect which can be interpreted as a tunneling phenomenon). \hl{However, QKF is valid only for spatially homogeneous fields. For linearly polarized fields, QKF consists of an integro-differential equation which is known in the literature as quantum Vlasov equation (QVE) \cite{kluger1998quantum, bloch1999pair, schmidt1999nonmarkov, dumlu2009equivalence, blaschke2009, blaschke2011, otto2015}. The general version of QKF extends the applicability to time-dependent electric fields of arbitrary polarizations \cite{aleksandrov2020generalQKE, aleksandrov2024generalQKE}.}
	
	Numerous studies have explored the influence of various field configurations on pair production. The impact of the laser frequency, carrier phase and pulse length on momentum distribution was studied for a short-pulse laser with subcycle structure in Ref.~\cite{hebenstreit2009momentum}. \hl{Other work has focused on the dynamically assisted Schwinger mechanism, which can significantly enhance the pair production rate \cite{schutzhold2008dynamically, nuriman2012enhanced, fey2012momentum, kohlfurst2013optimal, linder2015, li2021enhanced}.} Frequency chirping has been studied as another means to enhance the pair production rate \cite{dumlu2010schwinger, olugh2019pair, Min2013, gong2020electron}. It has also been observed that the temporal asymmetry in the electric field can enhance the particle density by a few orders of magnitude \cite{chen2024asymmetric, oluk2014electron, olugh2020asymmetric}. 
	
	In this work, we study the combined effects of the carrier-envelope phase and the temporal shape of the pulse on electron-positron pair production in the presence of a time-dependent asymmetric linearly polarized electric field. We numerically solve the QVE for three different bell-shaped envelopes. These envelopes are related to the Gaussian, Sauter-type and Lorentzian functions, and are defined to have a range of properties such as fast versus slow decay, flat-topped versus peaked. This approach allows us to present results that are broadly applicable to generic bell-shaped pulses.
	
	Our analysis shows that both the momentum distribution and the number density of produced pairs are highly sensitive to the carrier-envelope phase, pulse asymmetry and pulse shape. Using a semiclassical treatment based on the turning points of the potential of the equivalent 1D scattering problem in the complex-time plane, we confirm that the interference patterns in the momentum distribution arise from the interaction between these turning points. We also indicate the values of parameters for which the number density is particularly larger. Most plots in this work show results for the Gaussian envelope, as the differences between the three envelopes are often qualitatively similar; when this is not true, we present plots for all three. Throughout this study, we use natural units ($\hbar = c = 1$).
	
	The paper is organized as follows: In Sec.~\ref{sec:Theoretical formalism}, we briefly review the QVE and introduce the external electric field which we study in this work. Sec.~\ref{sec:Momentum spectra} presents numerical results for the momentum distribution, including a semiclassical analysis based on turning points. We discuss the number density in Sec.~\ref{sec:Number density} and conclude in Sec.~\ref{sec:Conclusion}.
	
	\section{Theoretical formalism : The quantum Vlasov equation}\label{sec:Theoretical formalism}
	
	\begin{figure*} [t]
		\includegraphics[width=\linewidth]{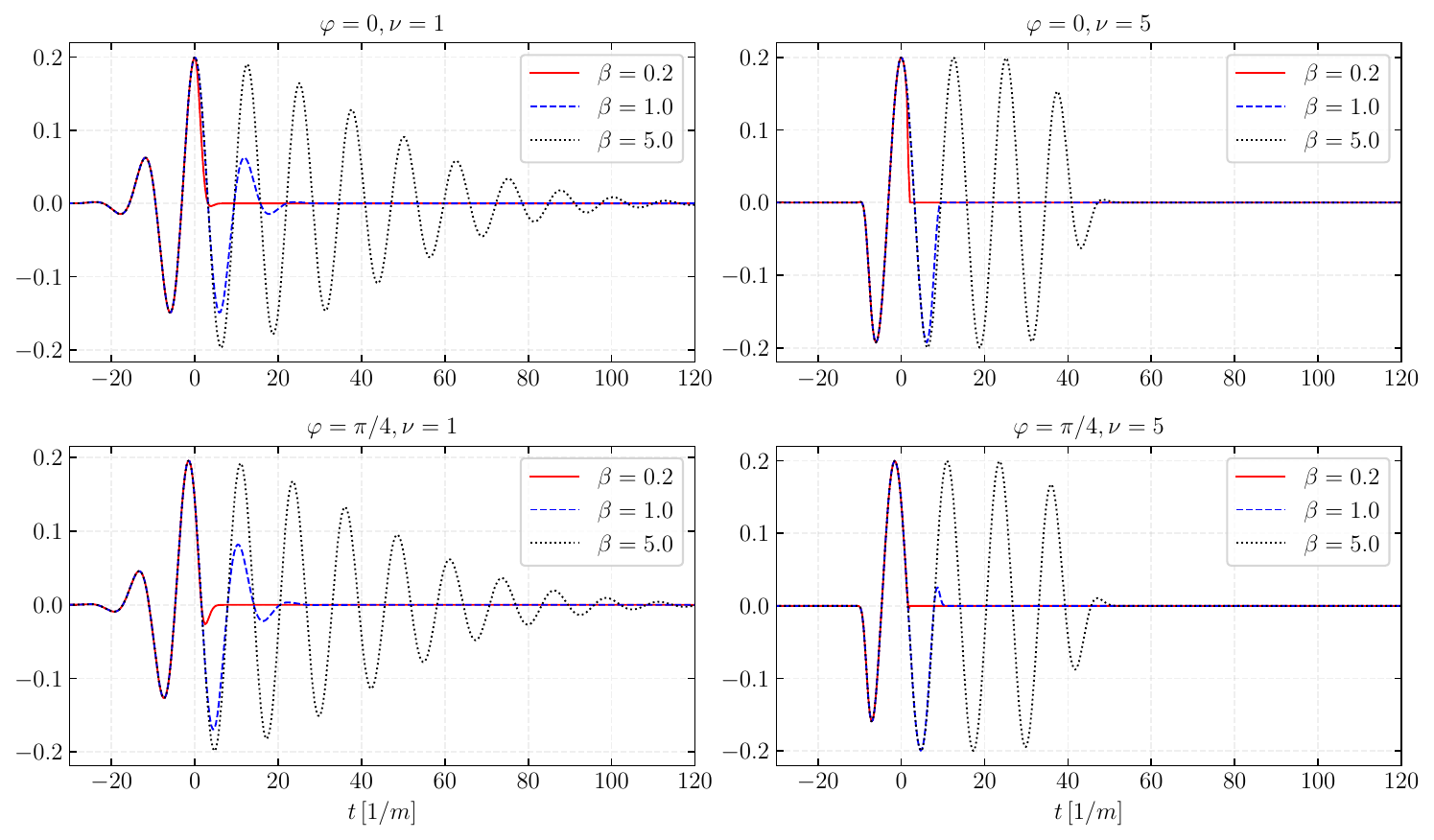}
		\caption{\small Electric field with a Gaussian envelope for several combinations of $\beta$, $\varphi$ and $\nu$.}
		\label{fig:field_profiles_gaussian}
	\end{figure*}
	
	\begin{figure*} [t]
		\includegraphics[width=\linewidth]{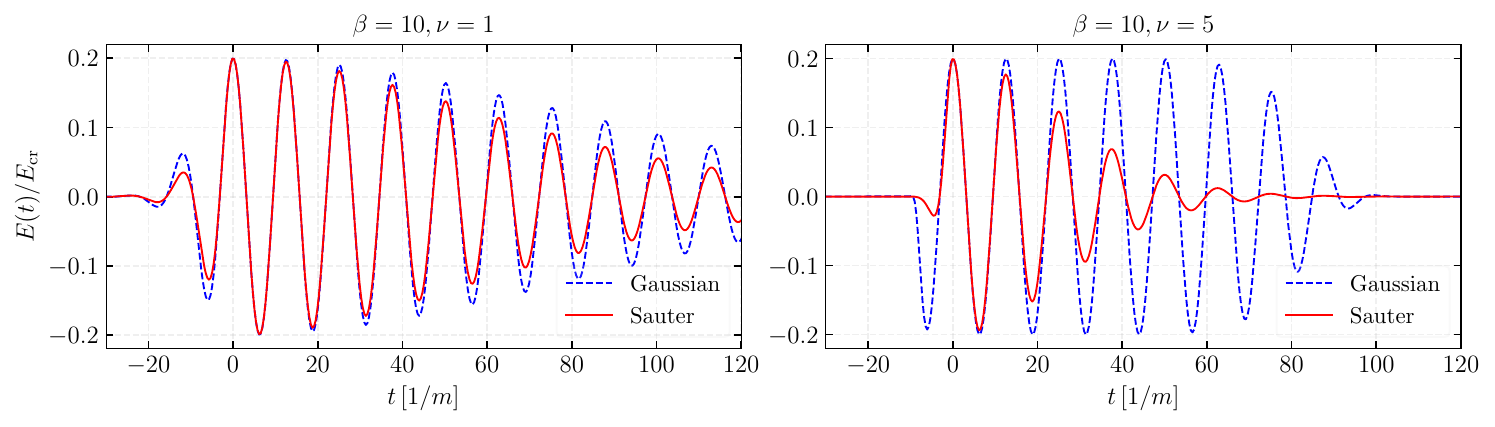}
		\caption{\small Comparison of electric field profiles with Gaussian (dashed blue) and Sauter (solid red) envelopes with a large asymmetry parameter ($\beta =10$). Unlike the Gaussian case, the Sauter envelope does not become flatter as $\nu$ is increased.}
		\label{fig:field_gaussian_vs_sauter}
	\end{figure*}

	\begin{figure*}[t]
		\centering
		\includegraphics[width=\textwidth]{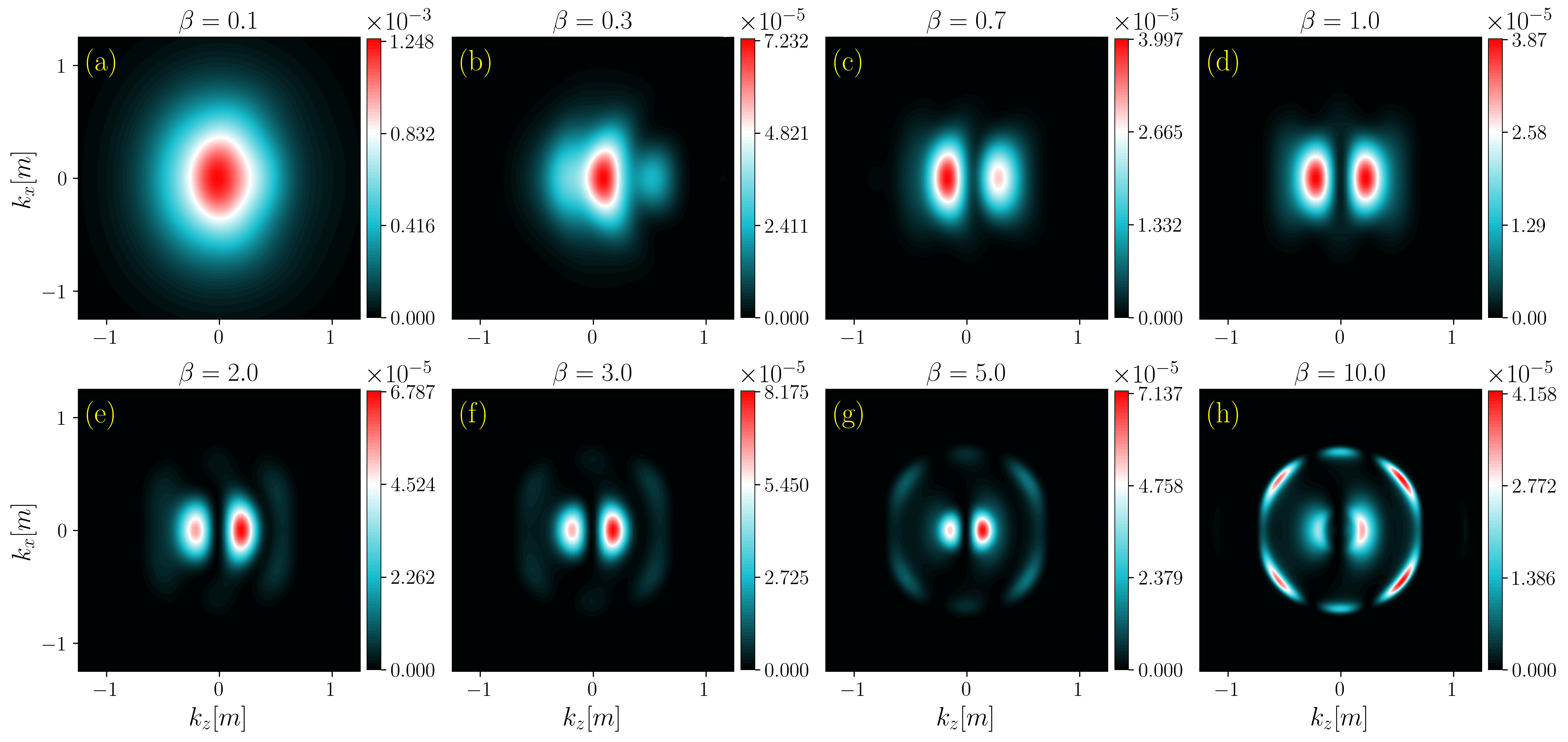}
		\caption{\small Momentum distribution $f_\textbf{k}(\infty)$ with $k_y = 0$ for a Gaussian envelope with $\varphi = 0$ and $\nu = 1$.}
		\label{fig:MD_betaall_phi0_nu1}
	\end{figure*}
	
	\begin{figure*}
		\centering
		\includegraphics[width=\textwidth]{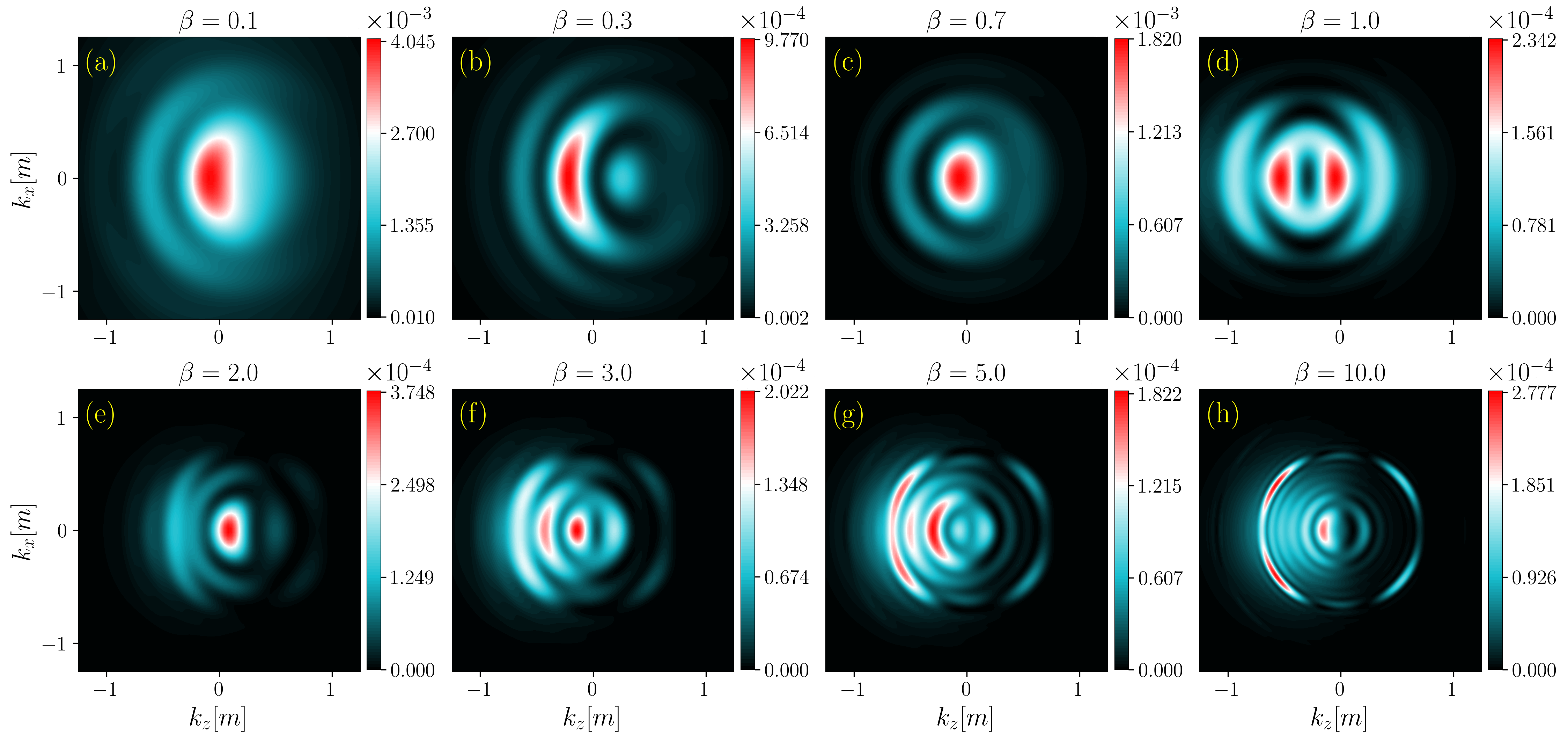}
		\caption{\small Same as Fig.~\ref{fig:MD_betaall_phi0_nu1} but with $\nu = 5$.}
		\label{fig:MD_betaall_phi0_nu5}
	\end{figure*}	
	
	The central quantity that we study in this work is the distribution function $f_\mathbf{k}(t)$. \hl{As the electric field that we have considered is linearly polarized and depends only on time, $f_\mathbf{k}(t)$ obeys the standard QVE \cite{kluger1998quantum, bloch1999pair, schmidt1999nonmarkov, dumlu2009equivalence, blaschke2009, blaschke2011, otto2015},} which is an integro-differential equation given by
	\begin{equation}
		\begin{split}\label{eq:QVE}
			\frac{df_\mathbf{k}(t)}{dt} & = \frac{1}{2} \, Q_\mathbf{k}(t) \int^t_{-\infty} dt' \, Q_\mathbf{k}(t') \, \left (1 - 2 f_\mathbf{k}(t') \right) \\
			& \qquad \qquad \times \,\cos \left(2\int^t_{t'}dt''\,\Omega_\mathbf{k}(t'')\right)
		\end{split}
	\end{equation}
	where
	\begin{equation}
		Q_\mathbf{k}(t) = \dfrac{e E(t) \,\sqrt{m^2 + \textbf{k}_\perp^2}}{\Omega^2_\mathbf{k}(t)},
	\end{equation}
	$\textbf{k} = (\textbf{k}_\perp, k_z)$ is the canonical momentum and 
	\begin{equation}\label{eq:dispersionRelation}
		\Omega_\mathbf{k}(t) = \sqrt{(k_z - eA(t))^2 + \textbf{k}_\perp^2 + m^2}
	\end{equation}
	is the instantaneous quasi-particle energy.
	
	\hl{At large times ($t\rightarrow\infty$), when the external electric field is turned off, the distribution function acquires a clear particle interpretation: $f_\mathbf{k}(\infty)$ represents the density of pairs produced, of which the positron has momentum $\mathbf{k}$ and the electron has momentum $-\mathbf{k}$}. A further requirement for such an interpretation is that the external electric field should vary slowly (adiabatic approximation) \cite{kluger1998quantum}. At intermediate times, however, the field is non-zero and $f_\mathbf{k}(t)$ contains contributions from both real and virtual excitations, so it does not admit a particle interpretation. \hl{Recently, in Ref.~\cite{ilderton2022interpretation}, it was shown that $f_\mathbf{k}(t)$ at an intermediate time $t$ has the following interpretation: it is the number of particles that would be measured (with a shifted momentum $\mathbf{k} - e \mathbf{A}(t)$) if the field were suddenly switched off at time $t$. As a sudden switch-off of an electric field is unphysical, $f_\mathbf{k}(t)$ at intermediate times remains unphysical. See Ref.~\cite{aleksandrov2025switchoff} for further analysis of this matter.} Due to these interpretational issues, we restrict our analysis to the asymptotic distribution function $f_\textbf{k}(\infty)$ and the total number of produced pairs per unit volume
	\begin{equation} \label{def:numberDensity}
		N = 2\int \frac{d^3\textbf{k}}{(2\pi)^3} f_\textbf{k}(\infty),
	\end{equation}
	where the factor of $2$ accounts for the spin degeneracy.
	
	The integro-differential QVE \eqref{eq:QVE} may be rewritten as a system of three coupled first-order ordinary differential equations \cite{blaschke2009}:
	\begin{equation}\label{eq:QVE_ODEs}
		\begin{split}
			\dfrac{df_\textbf{k}(t)}{dt} &= \dfrac{1}{2} \, Q_\textbf{k}(t)\, g_\textbf{k}(t),\\
			\dfrac{dg_\textbf{k}(t)}{dt} &= Q_\textbf{k}(t) \left(1 - 2f_\textbf{k}(t) \right) - 2 \, \Omega_\textbf{k}(t)\,h_\textbf{k}(t),\\
			\dfrac{dh_\textbf{k}(t)}{dt} &= 2 \, \Omega_\textbf{k}(t) \, g_\textbf{k}(t),
		\end{split}
	\end{equation}
	This system defines an initial-value problem with $f_\textbf{k}(-\infty) = g_\textbf{k}(-\infty) = h_\textbf{k}(-\infty) = 0$. This ODE formulation is considerably more convenient for numerical implementation than the original integro-differential equation \eqref{eq:QVE}. \hl{We integrate the system numerically using Runge-Kutta algorithms of various orders, including both adaptive and fixed-step schemes, over the interval $t = [- \eta \tau_1, \eta \tau_2]$, where $\tau_1$ and $\tau_2$ are the widths of the rising and falling pulses (see Sec.~\ref{subsec:fieldProfileDiscussion}), and $\eta > 0$ is chosen to be sufficiently large. The resulting solutions are insensitive to the choice of algorithm, integration step size, and to variations in $\eta$, confirming the stability and accuracy of our numerical procedure.}
	
	\subsection{External electric field configuration} \label{subsec:fieldProfileDiscussion}
	
	We consider a linearly polarized spatially homogeneous but time-dependent external electric field
	\begin{equation}
		\mathbf{E}(t) = E(t) \, \hat{z}.
	\end{equation}
	The corresponding vector potential $(A_0, \mathbf{A})$ may be chosen as
	\begin{equation}\label{def:vectorPotential}
		A_0 = 0, \quad \mathbf{A}(t) = A(t) \,\hat{z}, \quad \hl{A(t) = \int^{+\infty}_t E(t')\, dt'}.
	\end{equation}
	\hl{The choice of the reference point at $t=+\infty$ serves to remove the distinction between the canonical and kinetic momentum in the region of interest, which is the asymptotic region.}
	
	We consider time-dependent external electric fields of the form 
	\begin{equation}\label{eq:field_profile}
		E(t) = E_0 \cos(\omega t + \varphi) \Big(\mathcal{E}(t;\tau_1, \nu) \,\Theta(-t) + \mathcal{E}(t;\tau_2, \nu) \, \Theta(t)\Big)
	\end{equation}
	where $E_0$ is the amplitude, $\Theta(t)$ is the Heaviside step function, $\omega$ is the carrier frequency and $\varphi$ is the carrier-envelope phase (CEP). The object $\mathcal{E}(t;\tau, \nu)$ is a bell-shaped envelope centered at the origin, characterized by a width parameter $\tau$ and a steepness parameter $\nu$. The electric field becomes asymmetric about $t=0$ whenever $\tau_1 \neq \tau_2$ and, to quantify this asymmetry, we introduce the \textit{pulse asymmetry parameter} $\beta = \tau_2/\tau_1$. The falling edge of the pulse is compressed for $\beta < 1$ and elongated for $\beta > 1$. The order $\nu$ controls the steepness of the envelope on either side of $t=0$; for a super-Gaussian envelope $\nu$ is referred to as the \emph{super-Gaussian} order. \hl{A side-effect of using step functions is that higher derivatives of the field at $t=0$ are not continuous, even though the field itself is continuous. This has consequences for the turning-point analysis, which requires analytic continuation of the field to complex-time plane. We discuss these issues in Subsec.~\ref{subsec:TP} and Appx.~\ref{app:TP identification}.}
	 
	We have chosen three different bell-shaped envelopes in this work, in order to limit the dependence of our results on the finer details of the envelopes. \hl{These envelopes are normalized to unity at $t=0$} and are given by 
	\begin{align}\label{eq:envelope_Gaussian}
		& \text{Gaussian:} \qquad \mathcal{E}(t; \tau, \nu) = \exp \left(-\dfrac{t^{2\nu}}{2\tau^{2\nu}}\right)  \\ \label{eq:envelope_Lorentzian}
		& \text{Lorentzian:} \quad \mathcal{E}(t; \tau,\nu) = \dfrac{1}{1 + (t/\tau)^{2\nu}} \\ \label{eq:envelope_Sauter}
		& \text{Sauter:} \quad \quad \quad \mathcal{E}(t; \tau, \nu) = \text{sech}^{2\nu} \left(\dfrac{t}{\tau}\right)
	\end{align}
	\hl{We have studied composite fields \eqref{eq:field_profile} made up of the same kind of envelopes (e.g. Gaussian on both sides of $t=0$, but not Sauter on one side and Gaussian on the other).} Fig.~\ref{fig:field_profiles_gaussian} shows the electric field \eqref{eq:field_profile} with a Gaussian envelope for several values of $\beta$, $\varphi$ and $\nu$. A key distinction between the field profiles is that the Gaussian and Lorentzian envelopes become increasingly flat-topped as their width $\tau$ and the order $\nu$ are increased, whereas the Sauter envelope does not. The reason is simple: we have defined the Sauter envelope such that the power $2\nu$ acts on the function sech rather than its argument. Fig.~\ref{fig:field_gaussian_vs_sauter} illustrates this behavior. Whether or not an envelope develops a flat top has important implications for the momentum distribution of the produced pairs, as we will discuss in the next section.
	
	We now discuss the numerical values of the field parameters. For sub-critical fields ($E_0 < E_\text{cr}$) the density of produced pairs is low enough that the induced backreaction current can be neglected \cite{tanji2009dynamical, bloch1999pair, smolyansky2000, vinnik2001plasma, alkofer2001pair, roberts2002quantum, prakapenia2023pauli, jiang2023backreaction}. In this regime, the total electric field is essentially identical to the external electric field. Further, vacuum pair production by an external electric field can occur by two different mechanisms: a non-perturbative tunneling mechanism (the Schwinger effect) and a perturbative multiphoton absorption process. The Keldysh parameter $\gamma = m\omega/(e \, E_0)$ distinguishes between the two regimes: $\gamma \ll 1 $ indicates the non-perturbative regime, while $\gamma \gg 1$ indicates the perturbative regime \cite{keldysh1965ionization, brezin1970pair, popov2004Keldysh}. Throughout this work, we use the field parameters 
	\begin{equation} \label{def:fieldParameters}
		E_0 = 0.2\,E_\text{cr}, \quad \omega = 0.5\, m, \quad \tau_1 = 8/m.
	\end{equation}
	These values correspond to a subcritical field, which allows us to ignore backreaction effects. The Keldysh parameter evaluates to $\gamma = 2.5$, so we expect both perturbative and non-perturbative mechanisms to contribute to the pair production. \hl{It must be pointed out that the naive Keldysh parameter $\gamma = m\omega/(e \, E_0)$, which is technically only defined for monochromatic fields, may not adequately characterize the two regimes due to the presence of multiple scales in the problem at hand. Our results show that both perturbative and non-perturbative mechanisms contribute to the particle production, with varying degrees of dominance depending on the shape of the pulse. See Ref.~\cite{schutzhold2008dynamically} for an example of a modified Keldysh parameter in a multi-scale problem.}
	
	\hl{The homogeneous electric field \eqref{eq:field_profile} which we have discussed above should be considered an idealised theoretical setup, and we have chosen it in order to isolate specific behaviours of the system. If one views it as a \emph{dipole approximation} to a more realistic space-time dependent field, then the approximation is expected to be reasonable if the length scales in the electric field, such as the the envelope width in space and the carrier wavelength, are much larger than the pair creation length scale $l = m/(e E_0)$. The condition on carrier wavelenght is equivalent to $e E_0 \omega / m \ll 1$, which means large carrier frequencies invalidate the dipole approximation. For a discussion about the limitations of the dipole approximation we direct readers to Refs.~\cite{aleksandrov2017dipole, lv2018}. In these works it was shown that the relative error associated with the dipole approximation to rapidly oscillating fields can be several orders of magnitude. Interestingly, even in the expected range of validity of the approximation the relative error can be be around $100$\%.}
	
		\begin{figure*}[t]
		\includegraphics[width = 1.0\linewidth]{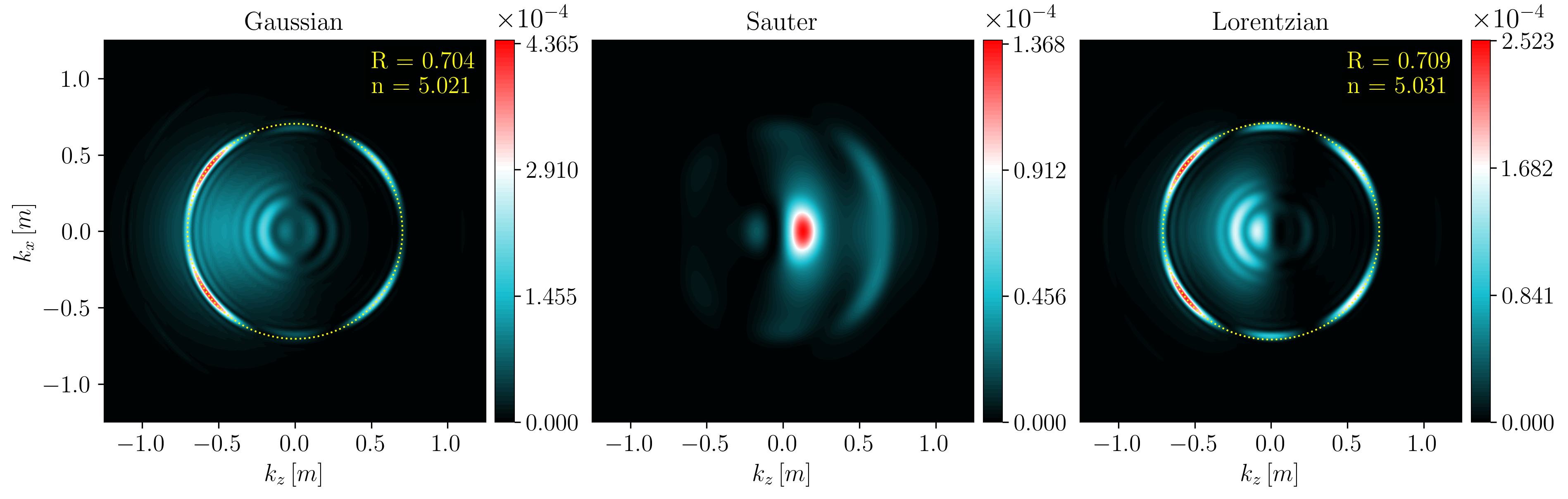}
		\caption{\small Ring-like structure in $f_\mathbf{k}(\infty)$ with $k_y = 0$ and $\varphi=0$ for a very long falling edge ($\beta=15$) and a large steepness parameter ($\nu=5$), characteristic of multiphoton pair production. The yellow overlaid circles mark the circular peaks associated with multiphoton absorption and their radii $R$ were determined numerically from the momentum distribution data. The corresponding photon number $n$, obtained from \eqref{eq:multiphoton} using $|\mathbf{k}| = R$, is not exactly an integer due to the finite pulse duration. No ring-like structure appears for the Sauter envelope as it does not acquire a flat-topped character at large $\nu$ and $\beta$.}
		\label{fig:multiphoton}
	\end{figure*}
	
	\begin{figure*}
		\centering
		\includegraphics[width = 1.0\linewidth]{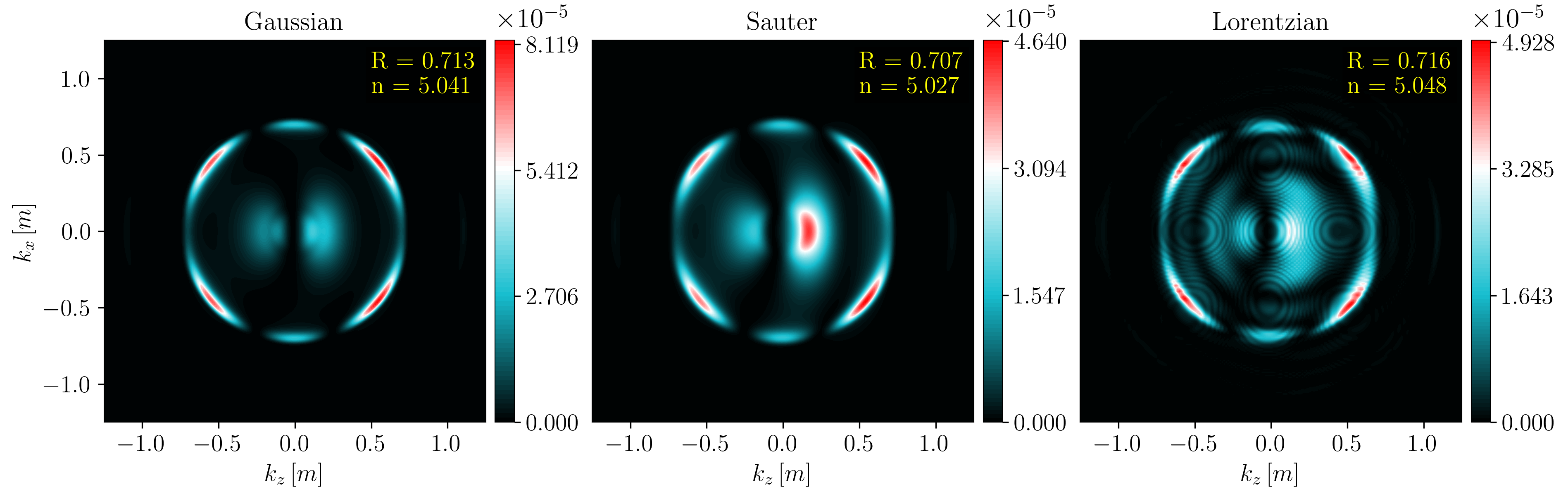}
		\caption{\small Same as Fig.~\ref{fig:multiphoton} but with $\nu=1$ and without the yellow overlaid circles. At this $\nu$, even the Sauter envelope produces a ring-like structure.}
		\label{fig:multiphoton2}
	\end{figure*}
	
	\section{Momentum distribution}\label{sec:Momentum spectra}

	In this section we present numerical results for the asymptotic momentum distribution of the produced pairs for a range of field parameters. As mentioned in the previous section, we fix the rising pulse width at $\tau_1 = 8/m$ and vary the falling pulse width with $\beta$ defined by $\tau_2 = \beta\tau_1$. 
	We also set $k_y = 0$ whenever plotting the momentum distributions. This does not entail a loss of generality because the system exhibits a cylindrical symmetry about the $z$-axis due to the fact that the electric field is linearly polarized and directed along that axis. Obviously, this cylindrical symmetry need not persist for other polarizations; \hl{see, e.g., Refs.~\cite{BlinnePhD, blinne2014rotating, olugh2019pair, aleksandrov2024rotating, li2015effects} and the references therein.} We remind the reader that the reference point of the vector potential \eqref{def:vectorPotential} was chosen at $t=+\infty$ so as to remove the distinction between the canonical and kinetic momentum in the asymptotic region.

	\subsection{Field asymmetry and steepness effects}
	
	In Fig.~\ref{fig:MD_betaall_phi0_nu1}, we show the asymptotic momentum distribution $f_\textbf{k}(\infty)$ for a Gaussian envelope with parameters $\nu = 1$ and $\varphi = 0$, illustrating the effects of a compressed falling pulse (top row) and an elongated falling pulse (bottom row). The panels are ordered such that the falling pulse width increases from top-left to bottom-right. As $\beta$ increases, the distribution undergoes a clear qualitative and quantitative change. 
	The peak value of the distribution falls sharply as $\beta$ increases, up to $\beta \approx 1$, and then increases gradually. The large peak for $\beta < 1 $ results from the strong asymmetry and \hl{steep \emph{switch-off} of the electric field \cite{adorno2015switchoff, linder2015, aleksandrov2025switchoff}}, whereas for $\beta > 1$ the more modest increase is driven solely by asymmetry. A detailed discussion of the effects of rapid switching on and off of strong electric backgrounds on pair production can be found in Ref.~\cite{aleksandrov2025switchoff}, where it is shown that sufficiently sharp field profiles can lead to an enhancement of pair production similar to that observed in the dynamically assisted Schwinger effect.
	
	In Fig.~\ref{fig:MD_betaall_phi0_nu5}, we again plot $f_\textbf{k}(\infty)$ for a (super-)Gaussian envelope with parameters $\nu = 5$ and $\varphi=0$. The parameter $\nu$ controls the steepness of the electric field (i.e. the switch-on switch-off rate) and, unlike $\beta$, does not alter the electric field's symmetry. Increasing $\beta$ at $\nu=5$ produces qualitative trends similar to those observed for $\nu=1$. However, for any fixed $\beta$, the distribution function now exhibits more peaks, and significantly taller ones, than in the $\nu=1$ case. This behavior is explained by the fact that increasing $\nu$ enhances the effective time during which the field remains near its maximum value. Another reason is the appearance of higher-frequency components in the Fourier spectrum of the electric field, which we discuss in greater detail in Sec.~\ref{sec:Number density}. Thus, our results clearly demonstrate that steepness has a greater effect on the particle production rate than asymmetry does. 
	
	We observe that the momentum distribution for the $\beta=1, \nu=1$ field (Fig.~\ref{fig:MD_betaall_phi0_nu1} (d)) exhibits an approximate $k_z$-reflection symmetry ($k_z \rightarrow -k_z$) while that for the $\beta=1, \nu=5$ field (Fig.~\ref{fig:MD_betaall_phi0_nu5} (d)) does not, despite both distributions corresponding to an electric field even in time. The net impulse $-e\int^\infty_{-\infty}E(t)\,dt$ carried by the $\beta=1, \nu=1$ field is negligible, allowing the vector potential to remain effectively odd and nearly preserving the symmetry. For the $\beta=1, \nu=5$ case the field profile is narrower, such that essentially only a few oscillation cycles fit in the pulse. Consequently, the cosine is no longer averaged out, and the net impulse has a non-negligible value resulting in the broken symmetry.
	
	We also observe that, for a fixed $\nu$, the number of peaks increases with $\beta$. This behavior is actually not universal and strongly depends on the field parameters. In Ref.~\cite{chen2024asymmetric} the number of peaks decreases as $\beta$ increases, whereas in Ref.~\cite{olugh2020asymmetric} it remains essentially constant as $\beta$ changes.
	
	For envelopes that become effectively flat-topped at large $\beta$, the outer peaks coalesce into ring-like interference structures. In fact, at large enough $\beta$ the ring-like structures can completely dominate the distribution; the central peaks only dominate if $\beta$ is small or if the envelope is not flat-topped. The rings also become more prominent as $\nu$ increases. These rings correspond to multiphoton absorption and emerge when the electric field features a long falling pulse \cite{BlinnePhD}. Their radius follows from energy conservation:
	\begin{equation}\label{eq:multiphoton}
		|\textbf{k}| = \sqrt{\left(\dfrac{n\omega}{2}\right)^2 - m_*^2},
	\end{equation}
	where $m_* = m\sqrt{1 + \dfrac{e^2 E_0^2}{2 m^2 \omega^2}}$ is the effective mass and $n$ denotes the number of absorbed photons \cite{kohlfurst2014effective}. Physically, this expression means that the absorption of $n$ photons of frequency $\omega$, shared between the electron and the positron, produces a characteristic excess momentum \textbf{k}. \hl{See Refs.~\cite{wollert2015, aleksandrov2018dase, aleksandrov2020spacetime, kohlfurst2024} for other applications of \eqref{eq:multiphoton} and its generalizations.} Although \eqref{eq:multiphoton} is strictly valid only for plane-wave fields, a faint partially formed ring is already visible at $\beta=3$ in Fig.~\ref{fig:MD_betaall_phi0_nu1} (f). The number of photons associated with the smallest ring follows directly from \eqref{eq:multiphoton}:
	\begin{equation}
		n_\text{min} = \text{ceil}\left(\dfrac{2 m_*}{\omega} \right).
	\end{equation}
	For the parameters used in this work ($E_0 = 0.2 E_\mathrm{cr}$, $\omega = 0.5\, m$), we obtain $n_\mathrm{min} = 5$.
	In Fig.~\ref{fig:multiphoton} (for $\nu = 5$) and in Fig.~\ref{fig:multiphoton2} (for $\nu = 1$) we plot $f_\textbf{k}(\infty)$ for a very long falling pulse $(\beta = 15)$. Multiphoton rings appear in all cases except for the $\nu=5$ Sauter envelope. As discussed above, the long or flat-topped falling pulse at $\beta = 15$ enhances the multiphoton contribution. This mechanism is absent from the Sauter envelope with $\nu=5$, which switches-off rapidly due to the large $\nu$ and does not produce rings. Fig.~\ref{fig:field_gaussian_vs_sauter} compares the Gaussian and Sauter envelopes with $\nu = 1$ and $\nu = 5$. \hl{See Ref.~\cite{aleksandrov2017shape} for a previous work comparing envelopes with and without flat-tops.} Also see Refs.~\cite{abdukerim2013effects, otto2018assisted, panferov2016assisted} for pair production by super-Gaussian envelopes.
	
	\subsection{Carrier-envelope phase effects}
	
	The carrier-envelope phase (CEP) $\varphi$ has a pronounced impact on the structure of the electric field profile, influencing both the positions of the extrema and the symmetry of the profile about $t=0$ (see Fig.~\ref{fig:field_profiles_gaussian}). It also determines how many oscillations of the carrier wave fit within the pulse envelope. Consequently, CEP plays an especially significant role for ultrashort pulses, where only a few carrier cycles lie within the envelope \cite{hebenstreit2009momentum, abdukerim2013effects}. \hl{See Ref.~\cite{dumlu2010stokes, aleksandrov2017shape, BraS2025} for more discussion on CEP effects in pair production. Also see \cite{kohlfurst2013optimal, hebestreit2014optimal} for use of optimal control theory to maximize the particle yield.}
	
	\begin{figure*}
		\centering
		\includegraphics[width=\textwidth]{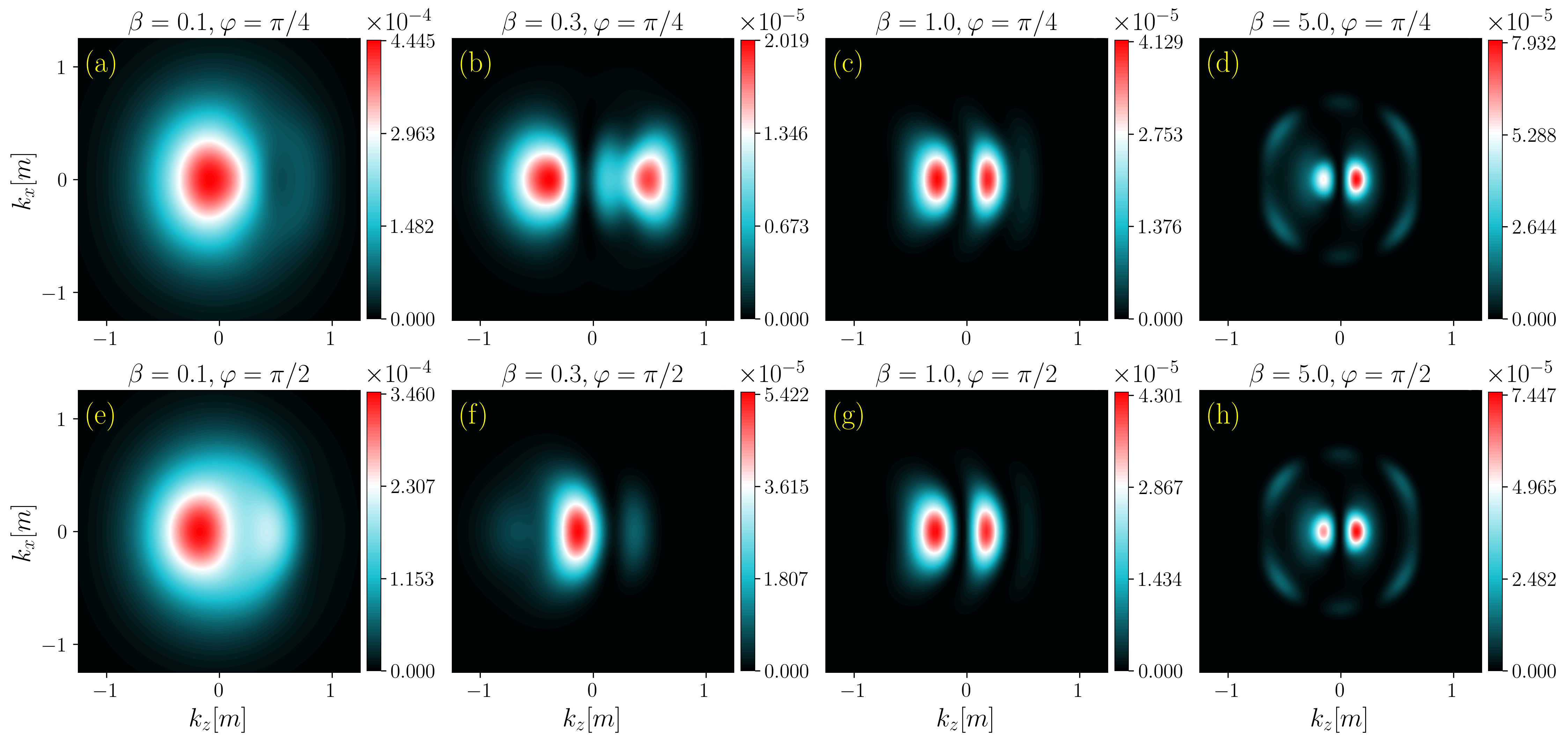}
		\caption{\small Momentum distribution $f_\textbf{k}(\infty)$ with $k_y=0$ for Gaussian envelope with non-zero $\varphi$ and $\nu=1$. Compare with Fig.~\ref{fig:MD_betaall_phi0_nu1}.}
		\label{fig:MD_beta_all_phi_nonzero_nu1}
	\end{figure*}
	
	\begin{figure*}
		\centering
		\includegraphics[width=\textwidth]{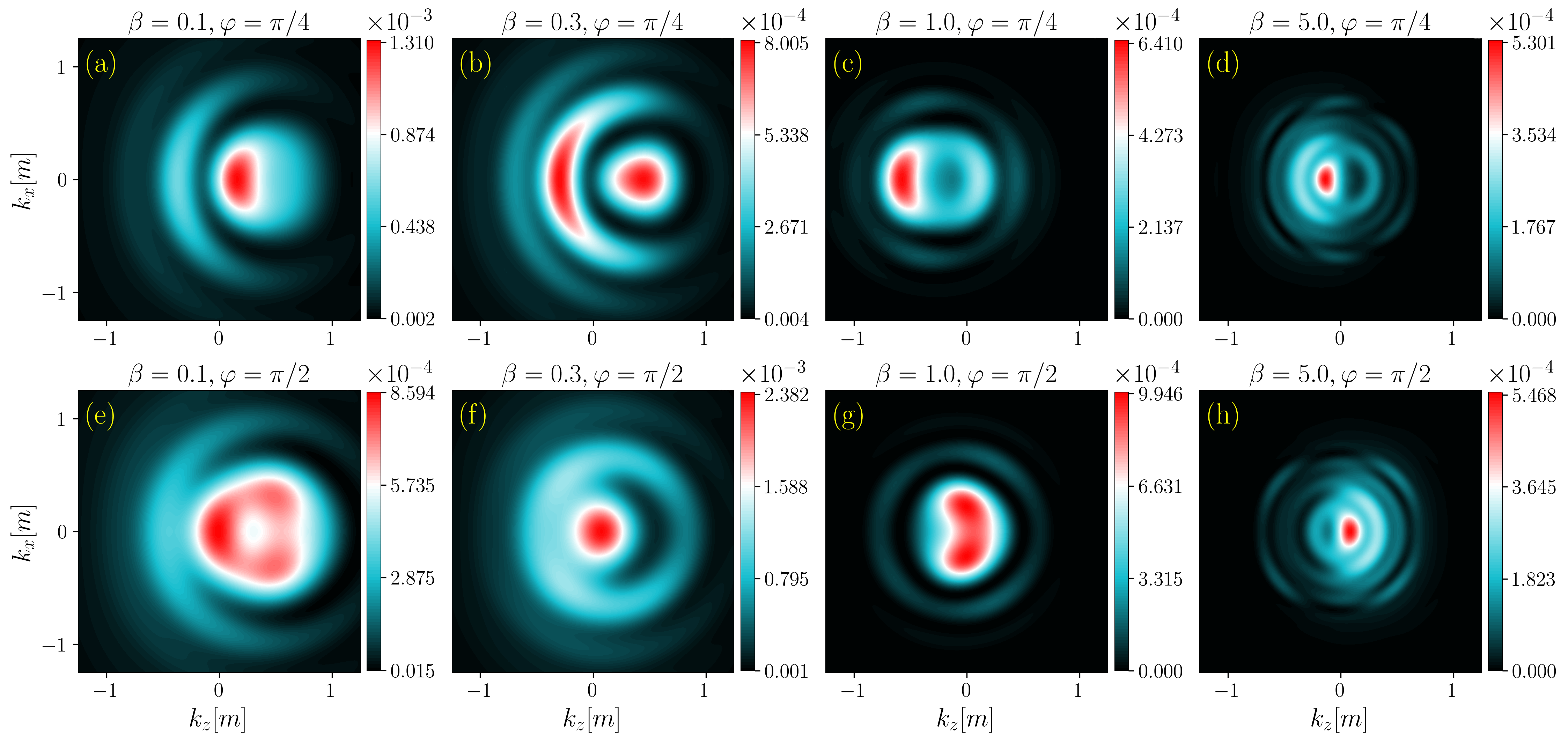}
		\caption{\small Same as Fig.~\ref{fig:MD_beta_all_phi_nonzero_nu1} but with $\nu = 5$. Compare with Fig.~\ref{fig:MD_betaall_phi0_nu5} }
		\label{fig:MD_beta_all_phi_nonzero_nu5}
	\end{figure*}
	
	Fig.~\ref{fig:MD_beta_all_phi_nonzero_nu1} (for $\nu=1$) and Fig.~\ref{fig:MD_beta_all_phi_nonzero_nu5} (for $\nu=5$) show the momentum distributions for a Gaussian envelope for various combinations of CEP ($\varphi$) and $\beta$. We find that the sensitivity of the momentum distribution and its peak to CEP, both in terms of its position and height, decreases as the width of the falling pulse increases. For very elongated falling pulses the effect is negligible, whereas for very short falling pulses a change in CEP can change the momentum distribution at and around the peak by an entire order of magnitude.
	
	Further comparison between Fig.~\ref{fig:MD_betaall_phi0_nu1} (d) and Fig.~\ref{fig:MD_beta_all_phi_nonzero_nu1} (c, g) shows that CEP alone can break the momentum-inversion symmetry. In distributions that already lack the inversion symmetry, CEP may exaggerate the asymmetry. In addition, the number of peaks in the momentum distribution is strongly influenced by CEP: Fig.~\ref{fig:MD_betaall_phi0_nu1} (b) exhibits a single dominant peak whereas for a nonzero CEP the distribution in Fig.~\ref{fig:MD_beta_all_phi_nonzero_nu1} (b) splits into two distinct peaks. See Refs.~\cite{mohamedsedik2023phase, mohamedsedik2021schwinger} for the effects of CEP on pair production in spatially inhomogeneous electric fields.
	
	\subsection{Semiclassical analysis} \label{subsec:TP}
	To interpret the features in the momentum distributions and number densities shown in Secs.~\ref{sec:Momentum spectra} and \ref{sec:Number density}, we employ a semiclassical turning point (saddle point) analysis. It is based on the fact that vacuum decay in a spatially homogeneous time-dependent electric field can be recast as a one dimensional over-the-barrier quantum mechanical scattering problem. The pair-production amplitude is controlled by the classical action evaluated on trajectories in the complex-time plane \cite{brezin1970pair, popov1972pair, dumlu2010schwinger, dumlu2011interference, akkermans2012ramsey, taya2021exact}.
	
	The turning points $t_i$ of the equivalent scattering potential are defined by $\Omega_\textbf{k}(t_i) = 0$. 
	By applying the standard WKB treatment, the asymptotic momentum distribution is found to be \cite{dumlu2011interference}
	\begin{equation}\label{eq:semiclassical}
		f_\textbf{k}(\infty) \approx \sum_{i} e^{-2K_\textbf{k}^i} + \sum_{i \neq j} 2\cos(2 \, \theta_\textbf{k}^{ij})(-1)^{i-j}e^{-K_\textbf{k}^i - K_\textbf{k}^{j}}
	\end{equation}
	where
	\begin{equation}
		K_\textbf{k}^i = \left|\int_{t_i^*}^{t_i}\Omega_\textbf{k}(t)\,dt\right|, \quad \hl{\theta_\textbf{k}^{ij} = \mathrm{Re}\, \int_{t_i}^{t_j}\Omega_\textbf{k}(t)\,dt.}
	\end{equation}
	This equation is accurate provided $f_\textbf{k}(\infty) \ll 1$. \hl{Note that the expression for $\theta_\textbf{k}^{ij}$ in the original work \cite{dumlu2011interference} contains an error.}
	
	Each turning point $t_i$ contributes separately an exponentially suppressed term $e^{-2K_\textbf{k}^i}$. Turning points farther from the real axis have larger imaginary parts, yielding larger $K_\textbf{k}^i$ and thus weaker contributions. Interference between pairs of turning points is contained in the second term in \eqref{eq:semiclassical} and is appreciable only when the involved turning points have comparable values of $K_\mathbf{k}^i$; otherwise the cross term is exponentially suppressed. A practical guideline is thus: turning points closest to the real axis tend to dominate the distribution, and interference is strongest between turning point pairs with similar distances from the real axis. Of course, depending on the structure of $A(t)$, other pairs of turning points can also interfere strongly. 
	
	\begin{figure}[tbh]
		\centering
		\includegraphics[width=1.0\linewidth]{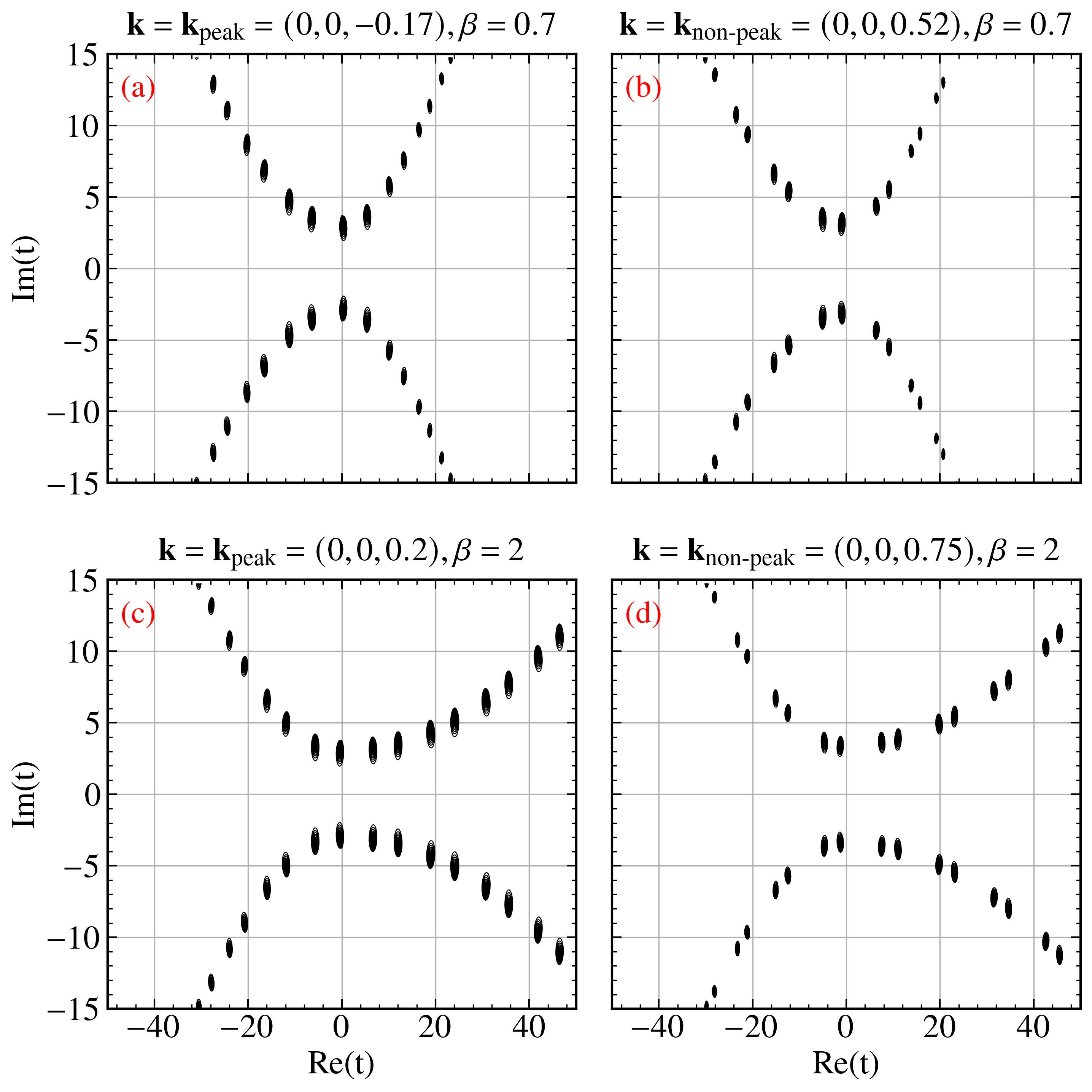}
		\caption{\small Contour plots of $|\Omega_\mathbf{k}(t)|^2$ in the complex-$t$ plane showing the locations of the turning points where $\Omega_\mathbf{k}(t) = 0$. These plots are for $\nu=1$ and $\varphi=0$. The figure illustrates the change in the turning points as the momentum is changed from the peak value (left panels) to a non-peak value (right panels) while keeping $\beta$ fixed.}
		\label{fig:TP_kz_variation}
	\end{figure}
	
	\hl{To study the turning point structure, we first need to analytically continue the field to the complex-time plane. However, since \eqref{eq:field_profile} is not smooth at $t=0$ due to the Heaviside step function, continuation to the complex-time plane is not possible. To enable a well-defined turning point analysis, we regularize the non-analyticity by replacing the Heaviside step function with a smooth step function that interpolates monotonically between $0$ and $1$ with slope controlled by a smoothing parameter. We have used smooth step functions of different analytic structures to ensure robustness of our results.} The details of this regularization and the numerical procedure used to identify the turning points are given in Appx.~\ref{app:TP identification}. In this section, we restrict our discussion to the results obtained for the Gaussian envelope.
	
	In Fig.~\ref{fig:TP_kz_variation}, panels (a) and (c) show the turning point structure at momenta corresponding to the peaks in Fig.~\ref{fig:MD_betaall_phi0_nu1} (c) and (d) respectively. Panels (b) and (d) in Fig.~\ref{fig:TP_kz_variation} show turning points at nearby momenta where the momentum distribution has a smaller value. In all cases the turning points arrange themselves in an infinitely tall hourglass structure symmetric about the real axis. The key difference lies in the proximity of turning points to the real axis: the momenta at the peaks correspond to turning points closer to the real axis as compared to those off the peaks. This aligns with the insight from \eqref{eq:semiclassical} and explains the larger values of the distribution at the peaks, and the weak interference in Figs.~\ref{fig:MD_betaall_phi0_nu1} (c) and (d) as most turning points beyond the first few are suppressed.
	
\begin{figure*}[tbh]
		\centering
		\includegraphics[width=1.0\linewidth]{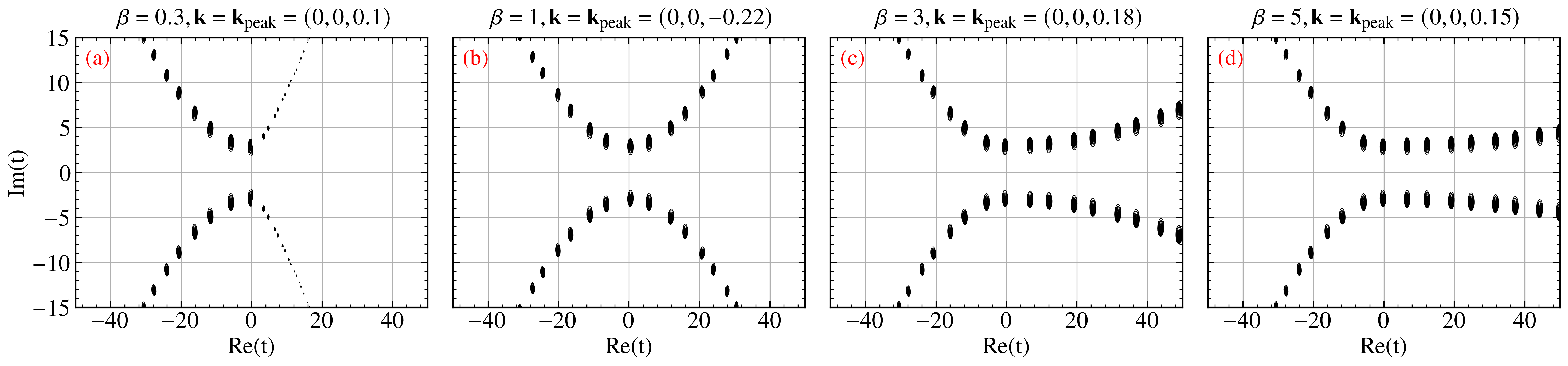}
		\caption{\small Contour plots of $|\Omega_\mathbf{k}(t)|^2$ in the complex-$t$ plane showing the locations of the turning points where $\Omega_\mathbf{k}(t) = 0$. These plots are for $\nu=1$ and $\varphi=0$ at peak momentum values for various $\beta$.}
		\label{fig:TP_beta_variation}
	\end{figure*}

	To illustrate how the asymmetry parameter $\beta$ modifies the interference structure, Fig.~\ref{fig:TP_beta_variation} shows turning points at peak momenta for several values of $\beta$. For small $\beta$, the hourglass shape is narrow and well-separated from the real axis. As $\beta$ increases, the hourglass broadens and the turning points move closer to the real axis. Consequently, a larger number of turning points contribute significantly to the second term in \eqref{eq:semiclassical}, leading to a pronounced strengthening of the interference patterns, as shown in Fig.~\ref{fig:MD_betaall_phi0_nu1}.
	
	\begin{figure}[t]
		\centering
		\includegraphics[width=1.0\linewidth]{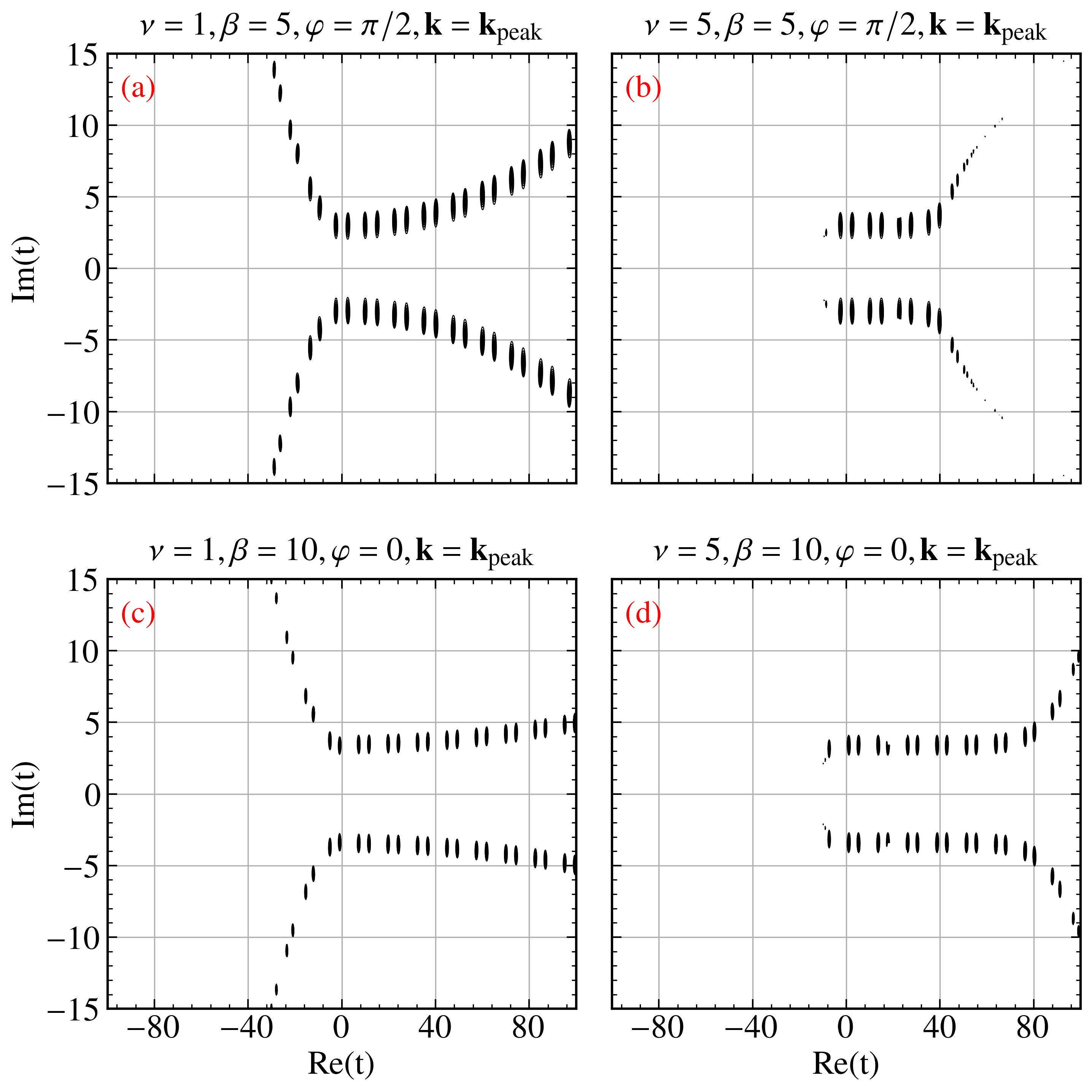}
		\caption{\small Contour plots of $|\Omega_\mathbf{k}(t)|^2$ in the complex-$t$ plane showing the locations of the turning points where $\Omega_\mathbf{k}(t) = 0$. The figure illustrates the change in the turning points as the field profile changes from Gaussian $\nu=1$ (left panels) to super-Gaussian $\nu=5$ (right panels). From upper left to lower right, $\mathbf{k}_\text{peak}$ are $(0,0,0.14)$, $(0,0,0.8)$, $(0.45,0,0.55)$, $(0.35, 0,-0.6)$.}
		\label{fig:TP_nu_variation}
	\end{figure}
	
	Fig.~\ref{fig:TP_nu_variation} illustrates how the super-Gaussian order $\nu$ affects the turning point structure. Panels (a) and (b) of Fig.~\ref{fig:TP_nu_variation} correspond to the peaks of Figs.~\ref{fig:MD_beta_all_phi_nonzero_nu1} (h) and \ref{fig:MD_beta_all_phi_nonzero_nu5} (h), respectively. Increasing $\nu$ brings more pairs of turning points to similar distances from the real axis, thereby raising the peak and contributing to a stronger interference pattern. A corresponding trend appears in panels (c) and (d) of  Fig.~\ref{fig:TP_nu_variation}, which explain the stronger interference and the larger peak value in Fig.~\ref{fig:MD_betaall_phi0_nu5} (h) compared to Fig.~\ref{fig:MD_betaall_phi0_nu1} (h). 
	
	\section{Number density}\label{sec:Number density}
\begin{figure*}[t]
		\includegraphics[width=\linewidth]{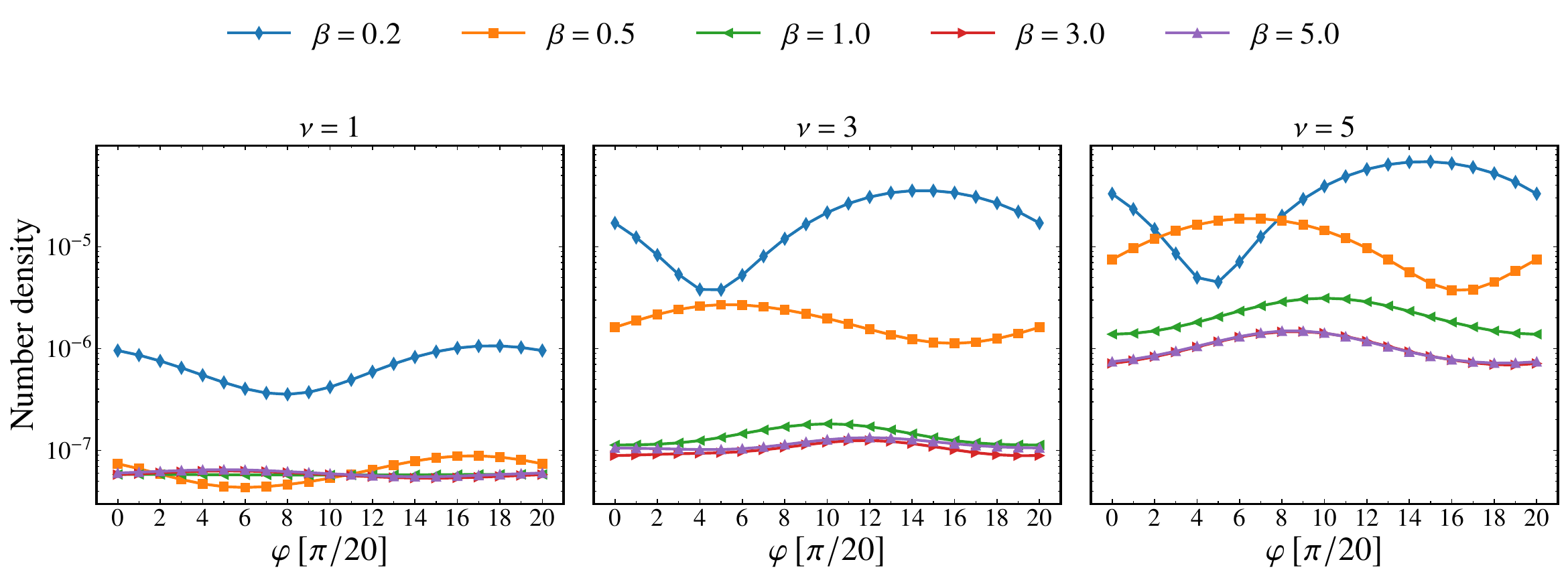}
		\caption{\small Number density (in units of $m^3$) as a function of the CEP $\varphi$ in (super-)Gaussian asymmetric electric fields.}
		\label{fig:number density}
	\end{figure*}

	In this section, we focus on the number density (short for total number of produced pairs per unit volume) as defined in Eq.~\eqref{def:numberDensity} in asymmetric (super-)Gaussian electric fields. See Fig.~\ref{fig:number density} for plots of the number density as a function of the CEP $\varphi$, for several values of the asymmetry parameter $\beta$ and the super-Gaussian order $\nu$. 
	
	It is clear from the plots that field profiles with shorter falling pulses (large $\nu$ or small $\beta$) consistently yield higher densities. The effect of steepness, which is controlled by $\nu$ and $\beta$ independently, outsizes that of asymmetry, which is controlled only by $\beta$. This aligns with the discussion in Sec.~\ref{sec:Momentum spectra}. Increasing $\nu$ increases the higher frequency components in the Fourier spectrum of the electric field, which leads to a significant enhancement in pair production due to the fact that higher frequency components can introduce multiphoton channels thereby enhancing pair production beyond the purely tunneling regime \cite{aleksandrov2025switchoff}. See Appx.~\ref{app:Fourier_field} for a discussion on the Fourier analysis of the electric field, and Fig.~\ref{fig:Fourier_analysis} for Fourier spectra of the electric field for various values of $\beta$ and $\nu$. On decreasing $\beta$, the two dominant peaks in the Fourier spectra become shorter and wider; thus $\beta$ also has a steepness effect, but less pronounced than that of $\nu$. As a result, we see an overall decrease in the number density with increase in $\beta$ values.
	
	Fig.~\ref{fig:number density} also shows that the sensitivity of the number density to the CEP $\varphi$ is greatly affected by $\nu$ and $\beta$. At $\nu=1$ the distribution is barely sensitive to CEP, unless $\beta$ is very small, whereas at $\nu=5$ and small $\beta$ a change in CEP can cause the total yield to increase by two to three orders of magnitude. This behavior can be explained by the fact that CEP controls the number of oscillations in the envelope much more strongly when the field has a short falling pulse (large $\nu$ or small $\beta$). 
	An elongated falling pulse does not preserve the subcycle structure and the long pulse width averages out the effect of CEP. A similar observation was made in Ref.~\cite{abdukerim2013effects} where the authors find the number density to have a higher sensitivity to $\varphi$ in the subcycle pulse case as compared to the supercycle case.
	
	\section{Conclusion and discussion}\label{sec:Conclusion}
	In this work, we have investigated the combined effects of carrier-envelope phase (CEP) and temporal shape of the pulse on pair production and momentum distribution in asymmetric electric fields with bell-shaped envelopes. We used three different bell-shaped envelopes (Gaussian, Sauter and Lorentzian) in order to be able to derive general results which do not depend on the finer details of the envelope. In particular, the Sauter envelope was defined to never be flat-topped in any limit. We used separate parameters to control the steepness and the asymmetry, so as to be able to differentiate their effects. The momentum distribution and the total number of produced pairs per unit volume were obtained by numerically solving the standard quantum Vlasov equation (QVE).
	
	The overall effect of asymmetry is to increase the peak values in the momentum distribution and, consquently, the number density of produced pairs. For asymmetric flat-topped field profiles with elongated falling pulses, we identified ring-shaped interference patterns corresponding to pair production by multiphoton absorption. The radii of the rings were found numerically, and they match very well with theoretical predictions based on simple energy conservation despite the fact that the field profiles are not plane waves. If the field profile is not flat-topped, then there are no significant multiphoton interference patterns.
	
	Decreasing the width of a field profile, which may be symmetric or not, increases the short-edge effect and enhances particle production. The short-edge effect is easily explained by the appearance of higher frequency peaks in the Fourier spectrum of the electric field.
	This enhancement is significantly stronger than that achieved by increasing the asymmetry in the field profile. Therefore, it is better to shorten the falling pulse in order to increase the pair production.
	
	We also found that CEP is a critical control parameter for short pulses as the momentum distribution and number density are extremely sensitive to CEP for short pulses. The effect of CEP diminishes as the pulse width increases. We found the number density to be the largest when the field profile was very short and had a non-zero CEP; however, this statement is sensitive to the field parameters that were kept fixed in this study (such as the electric field amplitude and carrier frequency).
	We have explained the interference patterns in the momentum distribution by analyzing the turning points of the single-particle dispersion relation. The turning points form an hourglass structure in the complex-time plane. If the hourglass is broader for some parameter values, then the turning points are closer to the real axis. This causes the interference effects to become stronger and the fringes more numerous.
	
	
	\appendix
	
	\section{Fourier Analysis}\label{app:Fourier_field}
\begin{figure*}[tbh]
		\includegraphics[width=\linewidth]{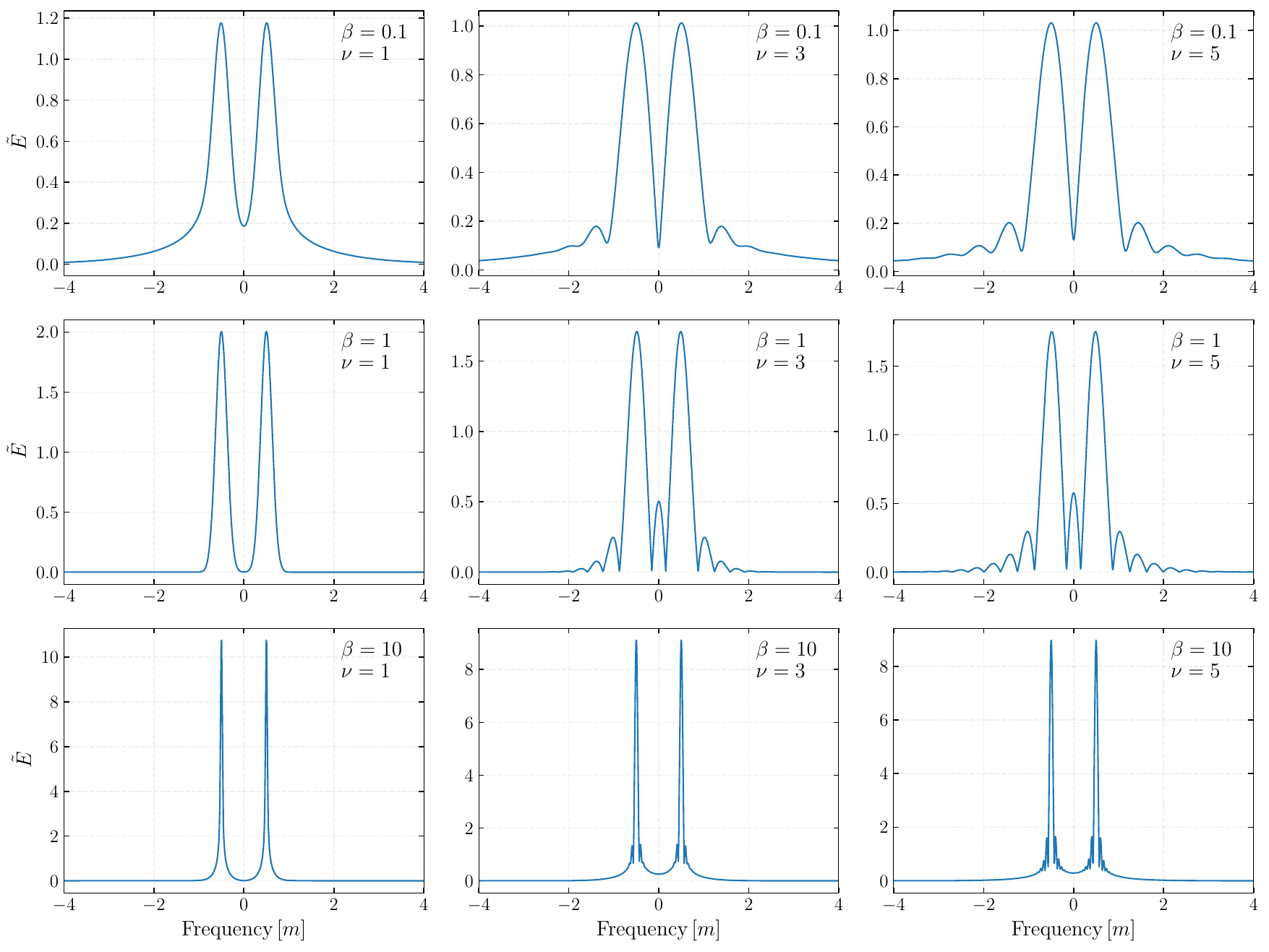}
		\caption{\small Fourier transform $\tilde{E}$ of the asymmetric (super-)Gaussian electric fields with $\varphi=0$. The two tallest peaks are located at $\omega'=\pm \omega$. Other peaks appear as $\nu$ is increased. The field parameters are the same as those used throughout this work: $E_0 = 0.2 E_\text{cr}, \omega = 0.5m$ and $\tau_1 = 8/m$.}
		\label{fig:Fourier_analysis}
	\end{figure*}

	In Fig.~\ref{fig:Fourier_analysis}, we plot the Fourier transform of the electric field
	\begin{equation}
		\tilde{E}(\omega') = \int^{\infty}_{-\infty} E(t)\,e^{i\omega't} dt. 
	\end{equation}
	as a function of the frequency $\omega'$ for (super-)Gaussian envelopes for various values of the asymmetry parameter $\beta$ and super-Gaussian order $\nu$. In each case, there are two dominant symmetric peaks. As $\nu$ increases, the envelope becomes steeper, leading to the appearance of higher frequency peaks in the Fourier spectra. These contributions, while having a relatively small weight, are important as they correspond to multiphoton channels that can substantially enhance the pair production. Consequently, increasing $\nu$ strengthens the high-frequency tail in the spectrum, which in turn boosts the overall number density. Also note that the peak is broader for smaller values of the pulse asymmetry parameter $\beta$. This explains the enhancement in particle production in the case of smaller values of $\beta$ (cf. Sec. \ref{sec:Number density}).
	
	\section{Numerical identification of turning points}\label{app:TP identification}	
	
	\hl{We regularize the non-analytic electric field by making the replacement
	\begin{equation}
		\Theta(t) \rightarrow \Theta_\epsilon(t) = \frac{1}{2} \left( 1 + g(t/\epsilon) \right), \quad \epsilon > 0
	\end{equation}
	where $g(t)$ is a symmetric smooth step function interpolating between $-1$ and $1$. The parameter $\epsilon$ serves as a smoothing parameter, controlling the slope at $t=0$. We have used $g(t) = \mathrm{erf} \,t$, $\mathrm{tanh}\, t$ and $t/\sqrt{1 + t^2}$. These functions differ significantly in their analytic properties: erf is an entire function, tanh has isolated poles along the imaginary axis, and $t/\sqrt{1 + t^2}$ possesses branch point singularities.}
	
	\hl{As $\Theta_\epsilon(t)$ reduces to $\Theta(t)$ in the limit $\epsilon \to 0$, we use small $\epsilon$ to plot turning points. We find that the locations of the turning points are independent of both the choice of $g(t)$ and value of $\epsilon$, with one caveat: we were unable to properly probe the erf-based step function for small $\epsilon$, as it becomes computationally expensive due to the rapid growth of erf along the imaginary direction. Nevertheless, even for larger $\epsilon$, the turning points obtained using erf remain robust and match precisely with those from the other step functions (verified up to $\epsilon = 5$). Thus we are confident that our regularization procedure is independent of the smooth step function and the smoothing parameter.}
	
	To locate the turning points of $\Omega_\mathbf{k}(t)$ in the complex-$t$ plane, we use an indirect contour-based method developed in Refs. \cite{dumlu2010schwinger, dumlu2011interference, dumlu2011complex}. Rather than solving $\Omega_\mathbf{k}(t) = 0$ directly, turning points are identified by plotting logarithmically spaced contours (on a uniformly spaced grid) defined by $|\Omega_\mathbf{k}(t)|^2 = \delta$
	with $\delta$ ranging from $10^{-10}$ to $10^{-1}$. The turning points manifest themselves as islands in these contour plots, as families of closed contours, parameterized by $\delta$, converge towards the turning points.


\begin{thebibliography}{86}%
		\makeatletter
		\providecommand \@ifxundefined [1]{%
			\@ifx{#1\undefined}
		}%
		\providecommand \@ifnum [1]{%
			\ifnum #1\expandafter \@firstoftwo
			\else \expandafter \@secondoftwo
			\fi
		}%
		\providecommand \@ifx [1]{%
			\ifx #1\expandafter \@firstoftwo
			\else \expandafter \@secondoftwo
			\fi
		}%
		\providecommand \natexlab [1]{#1}%
		\providecommand \enquote  [1]{``#1''}%
		\providecommand \bibnamefont  [1]{#1}%
		\providecommand \bibfnamefont [1]{#1}%
		\providecommand \citenamefont [1]{#1}%
		\providecommand \href@noop [0]{\@secondoftwo}%
		\providecommand \href [0]{\begingroup \@sanitize@url \@href}%
		\providecommand \@href[1]{\@@startlink{#1}\@@href}%
		\providecommand \@@href[1]{\endgroup#1\@@endlink}%
		\providecommand \@sanitize@url [0]{\catcode `\\12\catcode `\$12\catcode
			`\&12\catcode `\#12\catcode `\^12\catcode `\_12\catcode `\%12\relax}%
		\providecommand \@@startlink[1]{}%
		\providecommand \@@endlink[0]{}%
		\providecommand \url  [0]{\begingroup\@sanitize@url \@url }%
		\providecommand \@url [1]{\endgroup\@href {#1}{\urlprefix }}%
		\providecommand \urlprefix  [0]{URL }%
		\providecommand \Eprint [0]{\href }%
		\providecommand \doibase [0]{https://doi.org/}%
		\providecommand \selectlanguage [0]{\@gobble}%
		\providecommand \bibinfo  [0]{\@secondoftwo}%
		\providecommand \bibfield  [0]{\@secondoftwo}%
		\providecommand \translation [1]{[#1]}%
		\providecommand \BibitemOpen [0]{}%
		\providecommand \bibitemStop [0]{}%
		\providecommand \bibitemNoStop [0]{.\EOS\space}%
		\providecommand \EOS [0]{\spacefactor3000\relax}%
		\providecommand \BibitemShut  [1]{\csname bibitem#1\endcsname}%
		\let\auto@bib@innerbib\@empty
		\bibitem [{\citenamefont {Dirac}(1928)}]{dirac1928quantum}%
		\BibitemOpen
		\bibfield  {author} {\bibinfo {author} {\bibfnamefont {P.~A.~M.}\
				\bibnamefont {Dirac}},\ }\bibfield  {title} {\bibinfo {title} {The quantum
				theory of the electron},\ }\href {https://doi.org/10.1098/rspa.1928.0023}
		{\bibfield  {journal} {\bibinfo  {journal} {Proc. A 1}\ }\textbf {\bibinfo
				{volume} {117}},\ \bibinfo {pages} {610} (\bibinfo {year}
			{1928})}\BibitemShut {NoStop}%
		\bibitem [{\citenamefont {Sauter}(1931)}]{sauter1931behavior}%
		\BibitemOpen
		\bibfield  {author} {\bibinfo {author} {\bibfnamefont {F.}~\bibnamefont
				{Sauter}},\ }\bibfield  {title} {\bibinfo {title} {On the behavior of an
				electron in a homogeneous electric field in dirac’s relativistic theory},\
		}\href@noop {} {\bibfield  {journal} {\bibinfo  {journal} {Zeit. f. Phys}\
			}\textbf {\bibinfo {volume} {69}},\ \bibinfo {pages} {742} (\bibinfo {year}
			{1931})}\BibitemShut {NoStop}%
		\bibitem [{\citenamefont {Schwinger}(1951)}]{schwinger1951gauge}%
		\BibitemOpen
		\bibfield  {author} {\bibinfo {author} {\bibfnamefont {J.}~\bibnamefont
				{Schwinger}},\ }\bibfield  {title} {\bibinfo {title} {On gauge invariance and
				vacuum polarization},\ }\href {https://doi.org/10.1103/PhysRev.82.664}
		{\bibfield  {journal} {\bibinfo  {journal} {Phys. Rev.}\ }\textbf {\bibinfo
				{volume} {82}},\ \bibinfo {pages} {664} (\bibinfo {year} {1951})}\BibitemShut
		{NoStop}%
		\bibitem [{\citenamefont {Strickland}\ and\ \citenamefont
			{Mourou}(1985)}]{strickland1985compression}%
		\BibitemOpen
		\bibfield  {author} {\bibinfo {author} {\bibfnamefont {D.}~\bibnamefont
				{Strickland}}\ and\ \bibinfo {author} {\bibfnamefont {G.}~\bibnamefont
				{Mourou}},\ }\bibfield  {title} {\bibinfo {title} {Compression of amplified
				chirped optical pulses},\ }\href
		{https://doi.org/10.1016/0030-4018(85)90120-8} {\bibfield  {journal}
			{\bibinfo  {journal} {Optics communications}\ }\textbf {\bibinfo {volume}
				{55}},\ \bibinfo {pages} {447} (\bibinfo {year} {1985})}\BibitemShut
		{NoStop}%
		\bibitem [{\citenamefont {Yoon}\ \emph {et~al.}(2021)\citenamefont {Yoon} \emph
			{et~al.}}]{yoon2021realization}%
		\BibitemOpen
		\bibfield  {author} {\bibinfo {author} {\bibfnamefont {J.~W.}\ \bibnamefont
				{Yoon}} \emph {et~al.},\ }\bibfield  {title} {\bibinfo {title} {Realization
				of laser intensity over $10^{23}$ w/cm$^2$},\ }\href
		{https://doi.org/10.1364/OPTICA.420520} {\bibfield  {journal} {\bibinfo
				{journal} {Optica}\ }\textbf {\bibinfo {volume} {8}},\ \bibinfo {pages} {630}
			(\bibinfo {year} {2021})}\BibitemShut {NoStop}%
		\bibitem [{ELI()}]{ELI}%
		\BibitemOpen
		\href@noop {} {}\bibinfo {note} {\url{https://www.eli-laser.eu/}}\BibitemShut
		{NoStop}%
		\bibitem [{XFE()}]{XFEL}%
		\BibitemOpen
		\href@noop {} {}\bibinfo {note} {\url{https://www.xfel.eu/}}\BibitemShut
		{NoStop}%
		\bibitem [{\citenamefont {Ringwald}(2001)}]{ringwald2001pair}%
		\BibitemOpen
		\bibfield  {author} {\bibinfo {author} {\bibfnamefont {A.}~\bibnamefont
				{Ringwald}},\ }\bibfield  {title} {\bibinfo {title} {Pair production from
				vacuum at the focus of an x-ray free electron laser},\ }\href
		{https://doi.org/10.1016/S0370-2693(01)00496-8} {\bibfield  {journal}
			{\bibinfo  {journal} {Phys. Lett. B}\ }\textbf {\bibinfo {volume} {510}},\
			\bibinfo {pages} {107} (\bibinfo {year} {2001})}\BibitemShut {NoStop}%
		\bibitem [{\citenamefont {Khazanov}\ \emph {et~al.}(2023)\citenamefont
			{Khazanov} \emph {et~al.}}]{khazanov2023exawatt}%
		\BibitemOpen
		\bibfield  {author} {\bibinfo {author} {\bibfnamefont {E.}~\bibnamefont
				{Khazanov}} \emph {et~al.},\ }\bibfield  {title} {\bibinfo {title} {exawatt
				center for extreme light studies},\ }\href
		{https://doi.org/10.1017/hpl.2023.69} {\bibfield  {journal} {\bibinfo
				{journal} {High Power Laser Science and Engineering}\ }\textbf {\bibinfo
				{volume} {11}},\ \bibinfo {pages} {e78} (\bibinfo {year} {2023})}\BibitemShut
		{NoStop}%
		\bibitem [{\citenamefont {Gavrilov}\ and\ \citenamefont
			{Gitman}(1996)}]{gitman1996}%
		\BibitemOpen
		\bibfield  {author} {\bibinfo {author} {\bibfnamefont {S.~P.}\ \bibnamefont
				{Gavrilov}}\ and\ \bibinfo {author} {\bibfnamefont {D.~M.}\ \bibnamefont
				{Gitman}},\ }\bibfield  {title} {\bibinfo {title} {Vacuum instability in
				external fields},\ }\href {https://doi.org/10.1103/PhysRevD.53.7162}
		{\bibfield  {journal} {\bibinfo  {journal} {Phys. Rev. D}\ }\textbf {\bibinfo
				{volume} {53}},\ \bibinfo {pages} {7162} (\bibinfo {year}
			{1996})}\BibitemShut {NoStop}%
		\bibitem [{\citenamefont {Linder}\ \emph {et~al.}(2015)\citenamefont {Linder},
			\citenamefont {Schneider}, \citenamefont {Sicking}, \citenamefont {Szpak},\
			and\ \citenamefont {Sch\"utzhold}}]{linder2015}%
		\BibitemOpen
		\bibfield  {author} {\bibinfo {author} {\bibfnamefont {M.~F.}\ \bibnamefont
				{Linder}}, \bibinfo {author} {\bibfnamefont {C.}~\bibnamefont {Schneider}},
			\bibinfo {author} {\bibfnamefont {J.}~\bibnamefont {Sicking}}, \bibinfo
			{author} {\bibfnamefont {N.}~\bibnamefont {Szpak}},\ and\ \bibinfo {author}
			{\bibfnamefont {R.}~\bibnamefont {Sch\"utzhold}},\ }\bibfield  {title}
		{\bibinfo {title} {Pulse shape dependence in the dynamically assisted
				sauter-schwinger effect},\ }\href
		{https://doi.org/10.1103/PhysRevD.92.085009} {\bibfield  {journal} {\bibinfo
				{journal} {Phys. Rev. D}\ }\textbf {\bibinfo {volume} {92}},\ \bibinfo
			{pages} {085009} (\bibinfo {year} {2015})}\BibitemShut {NoStop}%
		\bibitem [{\citenamefont {W{\"o}llert}\ \emph {et~al.}(2015)\citenamefont
			{W{\"o}llert}, \citenamefont {Bauke},\ and\ \citenamefont
			{Keitel}}]{wollert2015}%
		\BibitemOpen
		\bibfield  {author} {\bibinfo {author} {\bibfnamefont {A.}~\bibnamefont
				{W{\"o}llert}}, \bibinfo {author} {\bibfnamefont {H.}~\bibnamefont {Bauke}},\
			and\ \bibinfo {author} {\bibfnamefont {C.~H.}\ \bibnamefont {Keitel}},\
		}\bibfield  {title} {\bibinfo {title} {Spin polarized electron-positron pair
				production via elliptical polarized laser fields},\ }\href
		{https://doi.org/10.1103/PhysRevD.91.125026} {\bibfield  {journal} {\bibinfo
				{journal} {Phys. Rev. D}\ }\textbf {\bibinfo {volume} {91}},\ \bibinfo
			{pages} {125026} (\bibinfo {year} {2015})}\BibitemShut {NoStop}%
		\bibitem [{\citenamefont {Aleksandrov}\ \emph {et~al.}(2016)\citenamefont
			{Aleksandrov}, \citenamefont {Plunien},\ and\ \citenamefont
			{Shabaev}}]{aleksandrov2016}%
		\BibitemOpen
		\bibfield  {author} {\bibinfo {author} {\bibfnamefont {I.~A.}\ \bibnamefont
				{Aleksandrov}}, \bibinfo {author} {\bibfnamefont {G.}~\bibnamefont
				{Plunien}},\ and\ \bibinfo {author} {\bibfnamefont {V.~M.}\ \bibnamefont
				{Shabaev}},\ }\bibfield  {title} {\bibinfo {title} {Electron-positron pair
				production in external electric fields varying both in space and time},\
		}\href {https://doi.org/10.1103/PhysRevD.94.065024} {\bibfield  {journal}
			{\bibinfo  {journal} {Phys. Rev. D}\ }\textbf {\bibinfo {volume} {94}},\
			\bibinfo {pages} {065024} (\bibinfo {year} {2016})}\BibitemShut {NoStop}%
		\bibitem [{\citenamefont {Lv}\ \emph {et~al.}(2018)\citenamefont {Lv} \emph
			{et~al.}}]{lv2018}%
		\BibitemOpen
		\bibfield  {author} {\bibinfo {author} {\bibfnamefont {Q.~Z.}\ \bibnamefont
				{Lv}} \emph {et~al.},\ }\bibfield  {title} {\bibinfo {title} {Role of the
				spatial inhomogeneity on the laser-induced vacuum decay},\ }\href
		{https://doi.org/10.1103/PhysRevA.97.022515} {\bibfield  {journal} {\bibinfo
				{journal} {Phys. Rev. A}\ }\textbf {\bibinfo {volume} {97}},\ \bibinfo
			{pages} {022515} (\bibinfo {year} {2018})}\BibitemShut {NoStop}%
		\bibitem [{\citenamefont {Popov}(1972)}]{popov1972pair}%
		\BibitemOpen
		\bibfield  {author} {\bibinfo {author} {\bibfnamefont {V.}~\bibnamefont
				{Popov}},\ }\bibfield  {title} {\bibinfo {title} {Pair production in a
				variable external field (quasiclassical approximation)},\ }\href@noop {}
		{\bibfield  {journal} {\bibinfo  {journal} {Soviet Journal of Experimental
					and Theoretical Physics}\ }\textbf {\bibinfo {volume} {34}},\ \bibinfo
			{pages} {709} (\bibinfo {year} {1972})}\BibitemShut {NoStop}%
		\bibitem [{\citenamefont {Di~Piazza}(2021)}]{dipiazza2021}%
		\BibitemOpen
		\bibfield  {author} {\bibinfo {author} {\bibfnamefont {A.}~\bibnamefont
				{Di~Piazza}},\ }\bibfield  {title} {\bibinfo {title} {Wkb electron wave
				functions in a tightly focused laser beam},\ }\href
		{https://doi.org/10.1103/PhysRevD.103.076011} {\bibfield  {journal} {\bibinfo
				{journal} {Phys. Rev. D}\ }\textbf {\bibinfo {volume} {103}},\ \bibinfo
			{pages} {076011} (\bibinfo {year} {2021})}\BibitemShut {NoStop}%
		\bibitem [{\citenamefont {Oertel}\ and\ \citenamefont
			{Sch\"utzhold}(2019)}]{oertel2019}%
		\BibitemOpen
		\bibfield  {author} {\bibinfo {author} {\bibfnamefont {J.}~\bibnamefont
				{Oertel}}\ and\ \bibinfo {author} {\bibfnamefont {R.}~\bibnamefont
				{Sch\"utzhold}},\ }\bibfield  {title} {\bibinfo {title} {Wkb approach to pair
				creation in spacetime-dependent fields: The case of a spacetime-dependent
				mass},\ }\href {https://doi.org/10.1103/PhysRevD.99.125014} {\bibfield
			{journal} {\bibinfo  {journal} {Phys. Rev. D}\ }\textbf {\bibinfo {volume}
				{99}},\ \bibinfo {pages} {125014} (\bibinfo {year} {2019})}\BibitemShut
		{NoStop}%
		\bibitem [{\citenamefont {Kohlf\"urst}\ \emph {et~al.}(2022)\citenamefont
			{Kohlf\"urst}, \citenamefont {Ahmadiniaz}, \citenamefont {Oertel},\ and\
			\citenamefont {Sch\"utzhold}}]{kohlfurst2019wkb}%
		\BibitemOpen
		\bibfield  {author} {\bibinfo {author} {\bibfnamefont {C.}~\bibnamefont
				{Kohlf\"urst}}, \bibinfo {author} {\bibfnamefont {N.}~\bibnamefont
				{Ahmadiniaz}}, \bibinfo {author} {\bibfnamefont {J.}~\bibnamefont {Oertel}},\
			and\ \bibinfo {author} {\bibfnamefont {R.}~\bibnamefont {Sch\"utzhold}},\
		}\bibfield  {title} {\bibinfo {title} {Sauter-schwinger effect for colliding
				laser pulses},\ }\href {https://doi.org/10.1103/PhysRevLett.129.241801}
		{\bibfield  {journal} {\bibinfo  {journal} {Phys. Rev. Lett.}\ }\textbf
			{\bibinfo {volume} {129}},\ \bibinfo {pages} {241801} (\bibinfo {year}
			{2022})}\BibitemShut {NoStop}%
		\bibitem [{\citenamefont {Dumlu}\ and\ \citenamefont
			{Dunne}(2010)}]{dumlu2010stokes}%
		\BibitemOpen
		\bibfield  {author} {\bibinfo {author} {\bibfnamefont {C.~K.}\ \bibnamefont
				{Dumlu}}\ and\ \bibinfo {author} {\bibfnamefont {G.~V.}\ \bibnamefont
				{Dunne}},\ }\bibfield  {title} {\bibinfo {title} {Stokes phenomenon and
				schwinger vacuum pair production in time-dependent laser pulses},\ }\href
		{https://doi.org/10.1103/PhysRevLett.104.250402} {\bibfield  {journal}
			{\bibinfo  {journal} {Phys. Rev. Lett.}\ }\textbf {\bibinfo {volume} {104}},\
			\bibinfo {pages} {250402} (\bibinfo {year} {2010})}\BibitemShut {NoStop}%
		\bibitem [{\citenamefont {Dunne}()}]{dunne2005fields}%
		\BibitemOpen
		\bibfield  {author} {\bibinfo {author} {\bibfnamefont {G.~V.}\ \bibnamefont
				{Dunne}},\ }\bibinfo {title} {Heisenberg--euler effective lagrangians: Basics
			and extensions},\ in\ \href {https://doi.org/10.1142/9789812775344_0014}
		{\emph {\bibinfo {booktitle} {From Fields to Strings: Circumnavigating
					Theoretical Physics}}},\ pp.\ \bibinfo {pages} {445--522}\BibitemShut
		{NoStop}%
		\bibitem [{\citenamefont {Dunne}\ and\ \citenamefont
			{Harris}(2023)}]{dunneharris2023}%
		\BibitemOpen
		\bibfield  {author} {\bibinfo {author} {\bibfnamefont {G.~V.}\ \bibnamefont
				{Dunne}}\ and\ \bibinfo {author} {\bibfnamefont {Z.~C.}\ \bibnamefont
				{Harris}},\ }\bibfield  {title} {\bibinfo {title} {Resurgence of the
				effective action in inhomogeneous fields},\ }\href
		{https://doi.org/10.1103/PhysRevD.107.065003} {\bibfield  {journal} {\bibinfo
				{journal} {Phys. Rev. D}\ }\textbf {\bibinfo {volume} {107}},\ \bibinfo
			{pages} {065003} (\bibinfo {year} {2023})}\BibitemShut {NoStop}%
		\bibitem [{\citenamefont {Gies}\ and\ \citenamefont
			{Karbstein}(2017)}]{gies2017}%
		\BibitemOpen
		\bibfield  {author} {\bibinfo {author} {\bibfnamefont {H.}~\bibnamefont
				{Gies}}\ and\ \bibinfo {author} {\bibfnamefont {F.}~\bibnamefont
				{Karbstein}},\ }\bibfield  {title} {\bibinfo {title} {An addendum to the
				heisenberg-euler effective action beyond one loop},\ }\href
		{https://doi.org/10.1007/JHEP03(2017)108} {\bibfield  {journal} {\bibinfo
				{journal} {J. High Energ. Phys}\ }\textbf {\bibinfo {volume} {2017}}
			(\bibinfo {year} {2017})}\BibitemShut {NoStop}%
		\bibitem [{\citenamefont {Dunne}\ and\ \citenamefont
			{Schubert}(2005)}]{dunne2005worldline}%
		\BibitemOpen
		\bibfield  {author} {\bibinfo {author} {\bibfnamefont {G.~V.}\ \bibnamefont
				{Dunne}}\ and\ \bibinfo {author} {\bibfnamefont {C.}~\bibnamefont
				{Schubert}},\ }\bibfield  {title} {\bibinfo {title} {Worldline instantons and
				pair production in inhomogenous fields},\ }\href
		{https://doi.org/10.1103/PhysRevD.72.105004} {\bibfield  {journal} {\bibinfo
				{journal} {Phys. Rev. D}\ }\textbf {\bibinfo {volume} {72}},\ \bibinfo
			{pages} {105004} (\bibinfo {year} {2005})}\BibitemShut {NoStop}%
		\bibitem [{\citenamefont {Ilderton}\ \emph {et~al.}(2015)\citenamefont
			{Ilderton}, \citenamefont {Torgrimsson},\ and\ \citenamefont
			{W\aa{}rdh}}]{ilderton2015worldline}%
		\BibitemOpen
		\bibfield  {author} {\bibinfo {author} {\bibfnamefont {A.}~\bibnamefont
				{Ilderton}}, \bibinfo {author} {\bibfnamefont {G.}~\bibnamefont
				{Torgrimsson}},\ and\ \bibinfo {author} {\bibfnamefont {J.}~\bibnamefont
				{W\aa{}rdh}},\ }\bibfield  {title} {\bibinfo {title} {Pair production from
				residues of complex worldline instantons},\ }\href
		{https://doi.org/10.1103/PhysRevD.92.025009} {\bibfield  {journal} {\bibinfo
				{journal} {Phys. Rev. D}\ }\textbf {\bibinfo {volume} {92}},\ \bibinfo
			{pages} {025009} (\bibinfo {year} {2015})}\BibitemShut {NoStop}%
		\bibitem [{\citenamefont {Olugh}\ \emph {et~al.}(2019)\citenamefont {Olugh},
			\citenamefont {Li}, \citenamefont {Xie},\ and\ \citenamefont
			{Alkofer}}]{olugh2019pair}%
		\BibitemOpen
		\bibfield  {author} {\bibinfo {author} {\bibfnamefont {O.}~\bibnamefont
				{Olugh}}, \bibinfo {author} {\bibfnamefont {Z.-L.}\ \bibnamefont {Li}},
			\bibinfo {author} {\bibfnamefont {B.-S.}\ \bibnamefont {Xie}},\ and\ \bibinfo
			{author} {\bibfnamefont {R.}~\bibnamefont {Alkofer}},\ }\bibfield  {title}
		{\bibinfo {title} {Pair production in differently polarized electric fields
				with frequency chirps},\ }\href {https://doi.org/10.1103/PhysRevD.99.036003}
		{\bibfield  {journal} {\bibinfo  {journal} {Phys. Rev. D}\ }\textbf {\bibinfo
				{volume} {99}},\ \bibinfo {pages} {036003} (\bibinfo {year}
			{2019})}\BibitemShut {NoStop}%
		\bibitem [{\citenamefont {Chen}\ \emph {et~al.}(2024)\citenamefont {Chen},
			\citenamefont {Amat}, \citenamefont {Hu}, \citenamefont {Fan},\ and\
			\citenamefont {Xie}}]{chen2024asymmetric}%
		\BibitemOpen
		\bibfield  {author} {\bibinfo {author} {\bibfnamefont {N.-Z.}\ \bibnamefont
				{Chen}}, \bibinfo {author} {\bibfnamefont {O.}~\bibnamefont {Amat}}, \bibinfo
			{author} {\bibfnamefont {L.-N.}\ \bibnamefont {Hu}}, \bibinfo {author}
			{\bibfnamefont {H.-H.}\ \bibnamefont {Fan}},\ and\ \bibinfo {author}
			{\bibfnamefont {B.-S.}\ \bibnamefont {Xie}},\ }\bibfield  {title} {\bibinfo
			{title} {Asymmetric pulse effects on pair production in chirped electric
				fields},\ }\href {https://doi.org/10.1103/PhysRevD.109.076015} {\bibfield
			{journal} {\bibinfo  {journal} {Phys. Rev. D}\ }\textbf {\bibinfo {volume}
				{109}},\ \bibinfo {pages} {076015} (\bibinfo {year} {2024})}\BibitemShut
		{NoStop}%
		\bibitem [{\citenamefont {Hebenstreit}\ \emph {et~al.}(2010)\citenamefont
			{Hebenstreit}, \citenamefont {Alkofer},\ and\ \citenamefont
			{Gies}}]{hebenstreit2010DHWvsQKE}%
		\BibitemOpen
		\bibfield  {author} {\bibinfo {author} {\bibfnamefont {F.}~\bibnamefont
				{Hebenstreit}}, \bibinfo {author} {\bibfnamefont {R.}~\bibnamefont
				{Alkofer}},\ and\ \bibinfo {author} {\bibfnamefont {H.}~\bibnamefont
				{Gies}},\ }\bibfield  {title} {\bibinfo {title} {Schwinger pair production in
				space- and time-dependent electric fields: Relating the wigner formalism to
				quantum kinetic theory},\ }\href {https://doi.org/10.1103/PhysRevD.82.105026}
		{\bibfield  {journal} {\bibinfo  {journal} {Phys. Rev. D}\ }\textbf {\bibinfo
				{volume} {82}},\ \bibinfo {pages} {105026} (\bibinfo {year}
			{2010})}\BibitemShut {NoStop}%
		\bibitem [{\citenamefont {Hebenstreit}(2011)}]{HebenstreitPhD}%
		\BibitemOpen
		\bibfield  {author} {\bibinfo {author} {\bibfnamefont {F.}~\bibnamefont
				{Hebenstreit}},\ }\emph {\bibinfo {title} {Schwinger effect in inhomogeneous
				electric fields}},\ \href@noop {} {Ph.D. thesis},\ \bibinfo  {school}
		{University of Graz} (\bibinfo {year} {2011})\BibitemShut {NoStop}%
		\bibitem [{\citenamefont {Blinne}(2016)}]{BlinnePhD}%
		\BibitemOpen
		\bibfield  {author} {\bibinfo {author} {\bibfnamefont {A.}~\bibnamefont
				{Blinne}},\ }\emph {\bibinfo {title} {Electron positron pair production in
				strong electric fields}},\ \href@noop {} {Ph.D. thesis},\ \bibinfo  {school}
		{University of Jena} (\bibinfo {year} {2016})\BibitemShut {NoStop}%
		\bibitem [{\citenamefont {Li}\ \emph {et~al.}(2015)\citenamefont {Li},
			\citenamefont {Lu},\ and\ \citenamefont {Xie}}]{li2015effects}%
		\BibitemOpen
		\bibfield  {author} {\bibinfo {author} {\bibfnamefont {Z.}~\bibnamefont
				{Li}}, \bibinfo {author} {\bibfnamefont {D.}~\bibnamefont {Lu}},\ and\
			\bibinfo {author} {\bibfnamefont {B.}~\bibnamefont {Xie}},\ }\bibfield
		{title} {\bibinfo {title} {Effects of electric field polarizations on pair
				production},\ }\href {https://doi.org/10.1103/PhysRevD.92.085001} {\bibfield
			{journal} {\bibinfo  {journal} {Phys. Rev. D}\ }\textbf {\bibinfo {volume}
				{92}},\ \bibinfo {pages} {085001} (\bibinfo {year} {2015})}\BibitemShut
		{NoStop}%
		\bibitem [{\citenamefont {Kluger}\ \emph {et~al.}(1998)\citenamefont {Kluger},
			\citenamefont {Mottola},\ and\ \citenamefont
			{Eisenberg}}]{kluger1998quantum}%
		\BibitemOpen
		\bibfield  {author} {\bibinfo {author} {\bibfnamefont {Y.}~\bibnamefont
				{Kluger}}, \bibinfo {author} {\bibfnamefont {E.}~\bibnamefont {Mottola}},\
			and\ \bibinfo {author} {\bibfnamefont {J.~M.}\ \bibnamefont {Eisenberg}},\
		}\bibfield  {title} {\bibinfo {title} {Quantum vlasov equation and its markov
				limit},\ }\href {https://doi.org/10.1103/PhysRevD.58.125015} {\bibfield
			{journal} {\bibinfo  {journal} {Phys. Rev. D}\ }\textbf {\bibinfo {volume}
				{58}},\ \bibinfo {pages} {125015} (\bibinfo {year} {1998})}\BibitemShut
		{NoStop}%
		\bibitem [{\citenamefont {Bloch}\ \emph {et~al.}(1999)\citenamefont {Bloch},
			\citenamefont {Mizerny}, \citenamefont {Prozorkevich}, \citenamefont
			{Roberts}, \citenamefont {Schmidt}, \citenamefont {Smolyansky},\ and\
			\citenamefont {Vinnik}}]{bloch1999pair}%
		\BibitemOpen
		\bibfield  {author} {\bibinfo {author} {\bibfnamefont {J.~C.}\ \bibnamefont
				{Bloch}}, \bibinfo {author} {\bibfnamefont {V.}~\bibnamefont {Mizerny}},
			\bibinfo {author} {\bibfnamefont {A.}~\bibnamefont {Prozorkevich}}, \bibinfo
			{author} {\bibfnamefont {C.~D.}\ \bibnamefont {Roberts}}, \bibinfo {author}
			{\bibfnamefont {S.}~\bibnamefont {Schmidt}}, \bibinfo {author} {\bibfnamefont
				{S.}~\bibnamefont {Smolyansky}},\ and\ \bibinfo {author} {\bibfnamefont
				{D.}~\bibnamefont {Vinnik}},\ }\bibfield  {title} {\bibinfo {title} {Pair
				creation: Back reactions and damping},\ }\href
		{https://doi.org/10.1103/PhysRevD.60.116011} {\bibfield  {journal} {\bibinfo
				{journal} {Phys. Rev. D}\ }\textbf {\bibinfo {volume} {60}},\ \bibinfo
			{pages} {116011} (\bibinfo {year} {1999})}\BibitemShut {NoStop}%
		\bibitem [{\citenamefont {Alkofer}\ \emph {et~al.}(2001)\citenamefont
			{Alkofer}, \citenamefont {Hecht}, \citenamefont {Roberts}, \citenamefont
			{Schmidt},\ and\ \citenamefont {Vinnik}}]{alkofer2001pair}%
		\BibitemOpen
		\bibfield  {author} {\bibinfo {author} {\bibfnamefont {R.}~\bibnamefont
				{Alkofer}}, \bibinfo {author} {\bibfnamefont {M.}~\bibnamefont {Hecht}},
			\bibinfo {author} {\bibfnamefont {C.~D.}\ \bibnamefont {Roberts}}, \bibinfo
			{author} {\bibfnamefont {S.}~\bibnamefont {Schmidt}},\ and\ \bibinfo {author}
			{\bibfnamefont {D.}~\bibnamefont {Vinnik}},\ }\bibfield  {title} {\bibinfo
			{title} {Pair creation and an x-ray free electron laser},\ }\href
		{https://doi.org/10.1103/PhysRevLett.87.193902} {\bibfield  {journal}
			{\bibinfo  {journal} {Phys. Rev. Lett.}\ }\textbf {\bibinfo {volume} {87}},\
			\bibinfo {pages} {193902} (\bibinfo {year} {2001})}\BibitemShut {NoStop}%
		\bibitem [{\citenamefont {Blaschke}\ \emph {et~al.}(2006)\citenamefont
			{Blaschke}, \citenamefont {Prozorkevich}, \citenamefont {Roberts},
			\citenamefont {Schmidt},\ and\ \citenamefont
			{Smolyansky}}]{blaschke2006pair}%
		\BibitemOpen
		\bibfield  {author} {\bibinfo {author} {\bibfnamefont {D.}~\bibnamefont
				{Blaschke}}, \bibinfo {author} {\bibfnamefont {A.}~\bibnamefont
				{Prozorkevich}}, \bibinfo {author} {\bibfnamefont {C.}~\bibnamefont
				{Roberts}}, \bibinfo {author} {\bibfnamefont {S.}~\bibnamefont {Schmidt}},\
			and\ \bibinfo {author} {\bibfnamefont {S.}~\bibnamefont {Smolyansky}},\
		}\bibfield  {title} {\bibinfo {title} {Pair production and optical lasers},\
		}\href {https://doi.org/10.1103/PhysRevLett.96.140402} {\bibfield  {journal}
			{\bibinfo  {journal} {Phys. Rev. Lett.}\ }\textbf {\bibinfo {volume} {96}},\
			\bibinfo {pages} {140402} (\bibinfo {year} {2006})}\BibitemShut {NoStop}%
		\bibitem [{\citenamefont {Dumlu}(2009)}]{dumlu2009equivalence}%
		\BibitemOpen
		\bibfield  {author} {\bibinfo {author} {\bibfnamefont {C.~K.}\ \bibnamefont
				{Dumlu}},\ }\bibfield  {title} {\bibinfo {title} {Quantum kinetic approach
				and the scattering approach to vacuum pair production},\ }\href
		{https://doi.org/10.1103/PhysRevD.79.065027} {\bibfield  {journal} {\bibinfo
				{journal} {Phys. Rev. D}\ }\textbf {\bibinfo {volume} {79}},\ \bibinfo
			{pages} {065027} (\bibinfo {year} {2009})}\BibitemShut {NoStop}%
		\bibitem [{\citenamefont {Orthaber}\ \emph {et~al.}(2011)\citenamefont
			{Orthaber}, \citenamefont {Hebenstreit},\ and\ \citenamefont
			{Alkofer}}]{orthaber2011momentum}%
		\BibitemOpen
		\bibfield  {author} {\bibinfo {author} {\bibfnamefont {M.}~\bibnamefont
				{Orthaber}}, \bibinfo {author} {\bibfnamefont {F.}~\bibnamefont
				{Hebenstreit}},\ and\ \bibinfo {author} {\bibfnamefont {R.}~\bibnamefont
				{Alkofer}},\ }\bibfield  {title} {\bibinfo {title} {Momentum spectra for
				dynamically assisted schwinger pair production},\ }\href
		{https://doi.org/10.1016/j.physletb.2011.02.053} {\bibfield  {journal}
			{\bibinfo  {journal} {Phys. Lett. B}\ }\textbf {\bibinfo {volume} {698}},\
			\bibinfo {pages} {80} (\bibinfo {year} {2011})}\BibitemShut {NoStop}%
		\bibitem [{\citenamefont {Aleksandrov}\ \emph {et~al.}(2020)\citenamefont
			{Aleksandrov}, \citenamefont {Dmitriev}, \citenamefont {Sevostyanov},\ and\
			\citenamefont {Smolyansky}}]{aleksandrov2020generalQKE}%
		\BibitemOpen
		\bibfield  {author} {\bibinfo {author} {\bibfnamefont {I.~A.}\ \bibnamefont
				{Aleksandrov}}, \bibinfo {author} {\bibfnamefont {V.~V.}\ \bibnamefont
				{Dmitriev}}, \bibinfo {author} {\bibfnamefont {D.~G.}\ \bibnamefont
				{Sevostyanov}},\ and\ \bibinfo {author} {\bibfnamefont {S.~A.}\ \bibnamefont
				{Smolyansky}},\ }\bibfield  {title} {\bibinfo {title} {Kinetic description of
				vacuum production in strong electric fields of arbitrary polarization},\
		}\href {https://doi.org/10.1140/epjst/e2020-000056-1} {\bibfield  {journal}
			{\bibinfo  {journal} {Eur. Phys. J. Spec. Top.}\ }\textbf {\bibinfo {volume}
				{229}},\ \bibinfo {pages} {3469} (\bibinfo {year} {2020})}\BibitemShut
		{NoStop}%
		\bibitem [{\citenamefont {Aleksandrov}\ \emph {et~al.}(2024)\citenamefont
			{Aleksandrov}, \citenamefont {Kudlis},\ and\ \citenamefont
			{Klochai}}]{aleksandrov2024generalQKE}%
		\BibitemOpen
		\bibfield  {author} {\bibinfo {author} {\bibfnamefont {I.~A.}\ \bibnamefont
				{Aleksandrov}}, \bibinfo {author} {\bibfnamefont {A.}~\bibnamefont
				{Kudlis}},\ and\ \bibinfo {author} {\bibfnamefont {A.~I.}\ \bibnamefont
				{Klochai}},\ }\bibfield  {title} {\bibinfo {title} {Kinetic theory of vacuum
				pair production in uniform electric fields revisited},\ }\href
		{https://doi.org/10.1103/PhysRevResearch.6.043009} {\bibfield  {journal}
			{\bibinfo  {journal} {Phys. Rev. Res.}\ }\textbf {\bibinfo {volume} {6}},\
			\bibinfo {pages} {043009} (\bibinfo {year} {2024})}\BibitemShut {NoStop}%
		\bibitem [{\citenamefont {Mocken}\ \emph {et~al.}(2010)\citenamefont {Mocken},
			\citenamefont {Ruf}, \citenamefont {M{\"u}ller},\ and\ \citenamefont
			{Keitel}}]{mocken2010nonperturbative}%
		\BibitemOpen
		\bibfield  {author} {\bibinfo {author} {\bibfnamefont {G.~R.}\ \bibnamefont
				{Mocken}}, \bibinfo {author} {\bibfnamefont {M.}~\bibnamefont {Ruf}},
			\bibinfo {author} {\bibfnamefont {C.}~\bibnamefont {M{\"u}ller}},\ and\
			\bibinfo {author} {\bibfnamefont {C.~H.}\ \bibnamefont {Keitel}},\ }\bibfield
		{title} {\bibinfo {title} {Nonperturbative multiphoton
				electron-positron--pair creation in laser fields},\ }\href
		{https://doi.org/10.1103/PhysRevA.81.022122} {\bibfield  {journal} {\bibinfo
				{journal} {Phys. Rev. A}\ }\textbf {\bibinfo {volume} {81}},\ \bibinfo
			{pages} {022122} (\bibinfo {year} {2010})}\BibitemShut {NoStop}%
		\bibitem [{\citenamefont {Ruf}\ \emph {et~al.}(2009)\citenamefont {Ruf},
			\citenamefont {Mocken}, \citenamefont {M{\"u}ller}, \citenamefont
			{Hatsagortsyan},\ and\ \citenamefont {Keitel}}]{ruf2009pair}%
		\BibitemOpen
		\bibfield  {author} {\bibinfo {author} {\bibfnamefont {M.}~\bibnamefont
				{Ruf}}, \bibinfo {author} {\bibfnamefont {G.~R.}\ \bibnamefont {Mocken}},
			\bibinfo {author} {\bibfnamefont {C.}~\bibnamefont {M{\"u}ller}}, \bibinfo
			{author} {\bibfnamefont {K.~Z.}\ \bibnamefont {Hatsagortsyan}},\ and\
			\bibinfo {author} {\bibfnamefont {C.~H.}\ \bibnamefont {Keitel}},\ }\bibfield
		{title} {\bibinfo {title} {Pair production in laser fields oscillating in
				space and time},\ }\href {https://doi.org/10.1103/PhysRevLett.102.080402}
		{\bibfield  {journal} {\bibinfo  {journal} {Phys. Rev. Lett.}\ }\textbf
			{\bibinfo {volume} {102}},\ \bibinfo {pages} {080402} (\bibinfo {year}
			{2009})}\BibitemShut {NoStop}%
		\bibitem [{\citenamefont {Kohlf{\"u}rst}\ \emph {et~al.}(2014)\citenamefont
			{Kohlf{\"u}rst}, \citenamefont {Gies},\ and\ \citenamefont
			{Alkofer}}]{kohlfurst2014effective}%
		\BibitemOpen
		\bibfield  {author} {\bibinfo {author} {\bibfnamefont {C.}~\bibnamefont
				{Kohlf{\"u}rst}}, \bibinfo {author} {\bibfnamefont {H.}~\bibnamefont
				{Gies}},\ and\ \bibinfo {author} {\bibfnamefont {R.}~\bibnamefont
				{Alkofer}},\ }\bibfield  {title} {\bibinfo {title} {Effective mass signatures
				in multiphoton pair production},\ }\href
		{https://doi.org/10.1103/PhysRevLett.112.050402} {\bibfield  {journal}
			{\bibinfo  {journal} {Phys. Rev. Lett.}\ }\textbf {\bibinfo {volume} {112}},\
			\bibinfo {pages} {050402} (\bibinfo {year} {2014})}\BibitemShut {NoStop}%
		\bibitem [{\citenamefont {Schmidt}\ \emph {et~al.}(1999)\citenamefont {Schmidt}
			\emph {et~al.}}]{schmidt1999nonmarkov}%
		\BibitemOpen
		\bibfield  {author} {\bibinfo {author} {\bibfnamefont {S.}~\bibnamefont
				{Schmidt}} \emph {et~al.},\ }\bibfield  {title} {\bibinfo {title}
			{Non-markovian effects in strong-field pair creation},\ }\href
		{https://doi.org/10.1103/PhysRevD.59.094005} {\bibfield  {journal} {\bibinfo
				{journal} {Phys. Rev. D}\ }\textbf {\bibinfo {volume} {59}},\ \bibinfo
			{pages} {094005} (\bibinfo {year} {1999})}\BibitemShut {NoStop}%
		\bibitem [{\citenamefont {Blaschke}\ \emph {et~al.}(2009)\citenamefont
			{Blaschke} \emph {et~al.}}]{blaschke2009}%
		\BibitemOpen
		\bibfield  {author} {\bibinfo {author} {\bibfnamefont {D.~B.}\ \bibnamefont
				{Blaschke}} \emph {et~al.},\ }\bibfield  {title} {\bibinfo {title} {Dynamical
				schwinger effect and high-intensity lasers. realising nonperturbative qed},\
		}\href {https://doi.org/10.1140/epjd/e2009-00156-y} {\bibfield  {journal}
			{\bibinfo  {journal} {Eur. Phys. J. D}\ }\textbf {\bibinfo {volume} {55}},\
			\bibinfo {pages} {341} (\bibinfo {year} {2009})}\BibitemShut {NoStop}%
		\bibitem [{\citenamefont {Blaschke}\ \emph {et~al.}(2011)\citenamefont
			{Blaschke}, \citenamefont {Dmitriev}, \citenamefont {R\"opke},\ and\
			\citenamefont {Smolyansky}}]{blaschke2011}%
		\BibitemOpen
		\bibfield  {author} {\bibinfo {author} {\bibfnamefont {D.~B.}\ \bibnamefont
				{Blaschke}}, \bibinfo {author} {\bibfnamefont {V.~V.}\ \bibnamefont
				{Dmitriev}}, \bibinfo {author} {\bibfnamefont {G.}~\bibnamefont {R\"opke}},\
			and\ \bibinfo {author} {\bibfnamefont {S.~A.}\ \bibnamefont {Smolyansky}},\
		}\bibfield  {title} {\bibinfo {title} {{BBGKY} kinetic approach for an
				${e}^{\ensuremath{-}}{e}^{+}\ensuremath{\gamma}$ plasma created from the
				vacuum in a strong laser-generated electric field: The one-photon
				annihilation channel},\ }\href {https://doi.org/10.1103/PhysRevD.84.085028}
		{\bibfield  {journal} {\bibinfo  {journal} {Phys. Rev. D}\ }\textbf {\bibinfo
				{volume} {84}},\ \bibinfo {pages} {085028} (\bibinfo {year}
			{2011})}\BibitemShut {NoStop}%
		\bibitem [{\citenamefont {Blaschke}\ \emph {et~al.}(2015)\citenamefont
			{Blaschke}, \citenamefont {Dmitriev}, \citenamefont {R\"opke},\ and\
			\citenamefont {Smolyansky}}]{otto2015}%
		\BibitemOpen
		\bibfield  {author} {\bibinfo {author} {\bibfnamefont {D.~B.}\ \bibnamefont
				{Blaschke}}, \bibinfo {author} {\bibfnamefont {V.~V.}\ \bibnamefont
				{Dmitriev}}, \bibinfo {author} {\bibfnamefont {G.}~\bibnamefont {R\"opke}},\
			and\ \bibinfo {author} {\bibfnamefont {S.~A.}\ \bibnamefont {Smolyansky}},\
		}\bibfield  {title} {\bibinfo {title} {Lifting shell structures in the
				dynamically assisted schwinger effect in periodic fields},\ }\href
		{https://doi.org/10.1016/j.physletb.2014.12.010} {\bibfield  {journal}
			{\bibinfo  {journal} {Phys. Lett. B}\ }\textbf {\bibinfo {volume} {740}},\
			\bibinfo {pages} {335} (\bibinfo {year} {2015})}\BibitemShut {NoStop}%
		\bibitem [{\citenamefont {Hebenstreit}\ \emph {et~al.}(2009)\citenamefont
			{Hebenstreit}, \citenamefont {Alkofer}, \citenamefont {Dunne},\ and\
			\citenamefont {Gies}}]{hebenstreit2009momentum}%
		\BibitemOpen
		\bibfield  {author} {\bibinfo {author} {\bibfnamefont {F.}~\bibnamefont
				{Hebenstreit}}, \bibinfo {author} {\bibfnamefont {R.}~\bibnamefont
				{Alkofer}}, \bibinfo {author} {\bibfnamefont {G.~V.}\ \bibnamefont {Dunne}},\
			and\ \bibinfo {author} {\bibfnamefont {H.}~\bibnamefont {Gies}},\ }\bibfield
		{title} {\bibinfo {title} {Momentum signatures for schwinger pair production
				in short laser pulses with a subcycle structure},\ }\href@noop {} {\bibfield
			{journal} {\bibinfo  {journal} {Phys. Rev. Lett.}\ }\textbf {\bibinfo
				{volume} {102}},\ \bibinfo {pages} {150404} (\bibinfo {year}
			{2009})}\BibitemShut {NoStop}%
		\bibitem [{\citenamefont {Sch{\"u}tzhold}\ \emph {et~al.}(2008)\citenamefont
			{Sch{\"u}tzhold}, \citenamefont {Gies},\ and\ \citenamefont
			{Dunne}}]{schutzhold2008dynamically}%
		\BibitemOpen
		\bibfield  {author} {\bibinfo {author} {\bibfnamefont {R.}~\bibnamefont
				{Sch{\"u}tzhold}}, \bibinfo {author} {\bibfnamefont {H.}~\bibnamefont
				{Gies}},\ and\ \bibinfo {author} {\bibfnamefont {G.}~\bibnamefont {Dunne}},\
		}\bibfield  {title} {\bibinfo {title} {Dynamically assisted schwinger
				mechanism},\ }\href {https://doi.org/10.1103/PhysRevLett.101.130404}
		{\bibfield  {journal} {\bibinfo  {journal} {Phys. Rev. Lett.}\ }\textbf
			{\bibinfo {volume} {101}},\ \bibinfo {pages} {130404} (\bibinfo {year}
			{2008})}\BibitemShut {NoStop}%
		\bibitem [{\citenamefont {Nuriman}\ \emph {et~al.}(2012)\citenamefont
			{Nuriman}, \citenamefont {Xie}, \citenamefont {Li},\ and\ \citenamefont
			{Sayipjamal}}]{nuriman2012enhanced}%
		\BibitemOpen
		\bibfield  {author} {\bibinfo {author} {\bibfnamefont {A.}~\bibnamefont
				{Nuriman}}, \bibinfo {author} {\bibfnamefont {B.-S.}\ \bibnamefont {Xie}},
			\bibinfo {author} {\bibfnamefont {Z.-L.}\ \bibnamefont {Li}},\ and\ \bibinfo
			{author} {\bibfnamefont {D.}~\bibnamefont {Sayipjamal}},\ }\bibfield  {title}
		{\bibinfo {title} {Enhanced electron--positron pair creation by dynamically
				assisted combinational fields},\ }\href
		{https://doi.org/10.1016/j.physletb.2012.09.060} {\bibfield  {journal}
			{\bibinfo  {journal} {Phys. Lett. B}\ }\textbf {\bibinfo {volume} {717}},\
			\bibinfo {pages} {465} (\bibinfo {year} {2012})}\BibitemShut {NoStop}%
		\bibitem [{\citenamefont {Fey}\ and\ \citenamefont
			{Sch{\"u}tzhold}(2012)}]{fey2012momentum}%
		\BibitemOpen
		\bibfield  {author} {\bibinfo {author} {\bibfnamefont {C.}~\bibnamefont
				{Fey}}\ and\ \bibinfo {author} {\bibfnamefont {R.}~\bibnamefont
				{Sch{\"u}tzhold}},\ }\bibfield  {title} {\bibinfo {title} {Momentum
				dependence in the dynamically assisted sauter-schwinger effect},\ }\href
		{https://doi.org/10.1103/PhysRevD.85.025004} {\bibfield  {journal} {\bibinfo
				{journal} {Phys. Rev. D}\ }\textbf {\bibinfo {volume} {85}},\ \bibinfo
			{pages} {025004} (\bibinfo {year} {2012})}\BibitemShut {NoStop}%
		\bibitem [{\citenamefont {Kohlf\"urst}\ \emph {et~al.}(2013)\citenamefont
			{Kohlf\"urst}, \citenamefont {Mitter}, \citenamefont {von Winckel},
			\citenamefont {Hebenstreit},\ and\ \citenamefont
			{Alkofer}}]{kohlfurst2013optimal}%
		\BibitemOpen
		\bibfield  {author} {\bibinfo {author} {\bibfnamefont {C.}~\bibnamefont
				{Kohlf\"urst}}, \bibinfo {author} {\bibfnamefont {M.}~\bibnamefont {Mitter}},
			\bibinfo {author} {\bibfnamefont {G.}~\bibnamefont {von Winckel}}, \bibinfo
			{author} {\bibfnamefont {F.}~\bibnamefont {Hebenstreit}},\ and\ \bibinfo
			{author} {\bibfnamefont {R.}~\bibnamefont {Alkofer}},\ }\bibfield  {title}
		{\bibinfo {title} {Optimizing the pulse shape for schwinger pair
				production},\ }\href {https://doi.org/10.1103/PhysRevD.88.045028} {\bibfield
			{journal} {\bibinfo  {journal} {Phys. Rev. D}\ }\textbf {\bibinfo {volume}
				{88}},\ \bibinfo {pages} {045028} (\bibinfo {year} {2013})}\BibitemShut
		{NoStop}%
		\bibitem [{\citenamefont {Li}\ \emph {et~al.}(2021)\citenamefont {Li},
			\citenamefont {Mohamedsedik},\ and\ \citenamefont {Xie}}]{li2021enhanced}%
		\BibitemOpen
		\bibfield  {author} {\bibinfo {author} {\bibfnamefont {L.-J.}\ \bibnamefont
				{Li}}, \bibinfo {author} {\bibfnamefont {M.}~\bibnamefont {Mohamedsedik}},\
			and\ \bibinfo {author} {\bibfnamefont {B.-S.}\ \bibnamefont {Xie}},\
		}\bibfield  {title} {\bibinfo {title} {Enhanced dynamically assisted pair
				production in spatial inhomogeneous electric fields with the frequency
				chirping},\ }\href {https://doi.org/10.1103/PhysRevD.104.036015} {\bibfield
			{journal} {\bibinfo  {journal} {Phys. Rev. D}\ }\textbf {\bibinfo {volume}
				{104}},\ \bibinfo {pages} {036015} (\bibinfo {year} {2021})}\BibitemShut
		{NoStop}%
		\bibitem [{\citenamefont {Dumlu}(2010)}]{dumlu2010schwinger}%
		\BibitemOpen
		\bibfield  {author} {\bibinfo {author} {\bibfnamefont {C.~K.}\ \bibnamefont
				{Dumlu}},\ }\bibfield  {title} {\bibinfo {title} {Schwinger vacuum pair
				production in chirped laser pulses},\ }\href
		{https://doi.org/10.1103/PhysRevD.82.045007} {\bibfield  {journal} {\bibinfo
				{journal} {Phys. Rev. D}\ }\textbf {\bibinfo {volume} {82}},\ \bibinfo
			{pages} {045007} (\bibinfo {year} {2010})}\BibitemShut {NoStop}%
		\bibitem [{\citenamefont {Min}\ \emph {et~al.}(2013)\citenamefont {Min} \emph
			{et~al.}}]{Min2013}%
		\BibitemOpen
		\bibfield  {author} {\bibinfo {author} {\bibfnamefont {J.}~\bibnamefont
				{Min}} \emph {et~al.},\ }\bibfield  {title} {\bibinfo {title} {Enhanced
				electron—positron pair creation by the frequency chirped laser pulse},\
		}\href {https://doi.org/10.1088/1674-1056/22/10/100307} {\bibfield  {journal}
			{\bibinfo  {journal} {Chin. Phys. B}\ }\textbf {\bibinfo {volume} {22}},\
			\bibinfo {pages} {100307} (\bibinfo {year} {2013})}\BibitemShut {NoStop}%
		\bibitem [{\citenamefont {Gong}\ \emph {et~al.}(2020)\citenamefont {Gong},
			\citenamefont {Li}, \citenamefont {Xie},\ and\ \citenamefont
			{Li}}]{gong2020electron}%
		\BibitemOpen
		\bibfield  {author} {\bibinfo {author} {\bibfnamefont {C.}~\bibnamefont
				{Gong}}, \bibinfo {author} {\bibfnamefont {Z.~L.}\ \bibnamefont {Li}},
			\bibinfo {author} {\bibfnamefont {B.~S.}\ \bibnamefont {Xie}},\ and\ \bibinfo
			{author} {\bibfnamefont {Y.~J.}\ \bibnamefont {Li}},\ }\bibfield  {title}
		{\bibinfo {title} {Electron-positron pair production in frequency modulated
				laser fields},\ }\href {https://doi.org/10.1103/PhysRevD.101.016008}
		{\bibfield  {journal} {\bibinfo  {journal} {Phys. Rev. D}\ }\textbf {\bibinfo
				{volume} {101}},\ \bibinfo {pages} {016008} (\bibinfo {year}
			{2020})}\BibitemShut {NoStop}%
		\bibitem [{\citenamefont {Oluk}\ \emph {et~al.}(2014)\citenamefont {Oluk},
			\citenamefont {Xie}, \citenamefont {Bake},\ and\ \citenamefont
			{Dulat}}]{oluk2014electron}%
		\BibitemOpen
		\bibfield  {author} {\bibinfo {author} {\bibfnamefont {O.}~\bibnamefont
				{Oluk}}, \bibinfo {author} {\bibfnamefont {B.-S.}\ \bibnamefont {Xie}},
			\bibinfo {author} {\bibfnamefont {M.~A.}\ \bibnamefont {Bake}},\ and\
			\bibinfo {author} {\bibfnamefont {S.}~\bibnamefont {Dulat}},\ }\bibfield
		{title} {\bibinfo {title} {Electron-positron pair production in a strong
				asymmetric laser electric field},\ }\href
		{https://doi.org/10.1007/s11467-013-0379-8} {\bibfield  {journal} {\bibinfo
				{journal} {Frontiers of Physics}\ }\textbf {\bibinfo {volume} {9}},\ \bibinfo
			{pages} {157} (\bibinfo {year} {2014})}\BibitemShut {NoStop}%
		\bibitem [{\citenamefont {Olugh}\ \emph {et~al.}(2020)\citenamefont {Olugh},
			\citenamefont {Li},\ and\ \citenamefont {Xie}}]{olugh2020asymmetric}%
		\BibitemOpen
		\bibfield  {author} {\bibinfo {author} {\bibfnamefont {O.}~\bibnamefont
				{Olugh}}, \bibinfo {author} {\bibfnamefont {Z.-L.}\ \bibnamefont {Li}},\ and\
			\bibinfo {author} {\bibfnamefont {B.-S.}\ \bibnamefont {Xie}},\ }\bibfield
		{title} {\bibinfo {title} {Asymmetric pulse effects on pair production in
				polarized electric fields},\ }\href {https://doi.org/10.1017/hpl.2020.36}
		{\bibfield  {journal} {\bibinfo  {journal} {High Power Laser Science and
					Engineering}\ }\textbf {\bibinfo {volume} {8}},\ \bibinfo {pages} {e38}
			(\bibinfo {year} {2020})}\BibitemShut {NoStop}%
		\bibitem [{\citenamefont {Ilderton}(2022)}]{ilderton2022interpretation}%
		\BibitemOpen
		\bibfield  {author} {\bibinfo {author} {\bibfnamefont {A.}~\bibnamefont
				{Ilderton}},\ }\bibfield  {title} {\bibinfo {title} {Physics of adiabatic
				particle number in the schwinger effect},\ }\href
		{https://doi.org/10.1103/PhysRevD.105.016021} {\bibfield  {journal} {\bibinfo
				{journal} {Phys. Rev. D}\ }\textbf {\bibinfo {volume} {105}},\ \bibinfo
			{pages} {016021} (\bibinfo {year} {2022})}\BibitemShut {NoStop}%
		\bibitem [{\citenamefont {Aleksandrov}\ \emph {et~al.}(2025)\citenamefont
			{Aleksandrov}, \citenamefont {Sevostyanov},\ and\ \citenamefont
			{Shabaev}}]{aleksandrov2025switchoff}%
		\BibitemOpen
		\bibfield  {author} {\bibinfo {author} {\bibfnamefont {I.~A.}\ \bibnamefont
				{Aleksandrov}}, \bibinfo {author} {\bibfnamefont {D.~G.}\ \bibnamefont
				{Sevostyanov}},\ and\ \bibinfo {author} {\bibfnamefont {V.~M.}\ \bibnamefont
				{Shabaev}},\ }\bibfield  {title} {\bibinfo {title} {Schwinger particle
				production: Rapid switch off of the external field versus dynamical
				assistance},\ }\href {https://doi.org/10.1103/PhysRevD.111.016010} {\bibfield
			{journal} {\bibinfo  {journal} {Phys. Rev. D}\ }\textbf {\bibinfo {volume}
				{111}},\ \bibinfo {pages} {016010} (\bibinfo {year} {2025})}\BibitemShut
		{NoStop}%
		\bibitem [{\citenamefont {Tanji}(2009)}]{tanji2009dynamical}%
		\BibitemOpen
		\bibfield  {author} {\bibinfo {author} {\bibfnamefont {N.}~\bibnamefont
				{Tanji}},\ }\bibfield  {title} {\bibinfo {title} {Dynamical view of pair
				creation in uniform electric and magnetic fields},\ }\href
		{https://doi.org/10.1016/j.aop.2009.03.012} {\bibfield  {journal} {\bibinfo
				{journal} {Annals of Physics}\ }\textbf {\bibinfo {volume} {324}},\ \bibinfo
			{pages} {1691} (\bibinfo {year} {2009})}\BibitemShut {NoStop}%
		\bibitem [{\citenamefont {Smolyansky}\ \emph {et~al.}(2000)\citenamefont
			{Smolyansky}, \citenamefont {Mizerny}, \citenamefont {Vinnik}, \citenamefont
			{Prozorkevich},\ and\ \citenamefont {Toneev}}]{smolyansky2000}%
		\BibitemOpen
		\bibfield  {author} {\bibinfo {author} {\bibfnamefont {S.}~\bibnamefont
				{Smolyansky}}, \bibinfo {author} {\bibfnamefont {V.}~\bibnamefont {Mizerny}},
			\bibinfo {author} {\bibfnamefont {D.}~\bibnamefont {Vinnik}}, \bibinfo
			{author} {\bibfnamefont {A.}~\bibnamefont {Prozorkevich}},\ and\ \bibinfo
			{author} {\bibfnamefont {V.}~\bibnamefont {Toneev}},\ }\bibfield  {title}
		{\bibinfo {title} {The non-equilibrium distribution function of particles and
				anti-particles created in strong fields},\ }in\ \href
		{https://doi.org/10.1142/9789812793812_0029} {\emph {\bibinfo {booktitle}
				{Progress in Nonequilibrium Green's Functions}}}\ (\bibinfo  {publisher}
		{World Scientific},\ \bibinfo {year} {2000})\ pp.\ \bibinfo {pages}
		{375--382}\BibitemShut {NoStop}%
		\bibitem [{\citenamefont {Vinnik}\ \emph {et~al.}(2001)\citenamefont {Vinnik}
			\emph {et~al.}}]{vinnik2001plasma}%
		\BibitemOpen
		\bibfield  {author} {\bibinfo {author} {\bibfnamefont {D.~V.}\ \bibnamefont
				{Vinnik}} \emph {et~al.},\ }\bibfield  {title} {\bibinfo {title} {Plasma
				production and thermalisation in a strong field},\ }\href
		{https://doi.org/10.1007/s100520100787} {\bibfield  {journal} {\bibinfo
				{journal} {Eur. Phys. J. C}\ }\textbf {\bibinfo {volume} {22}},\ \bibinfo
			{pages} {341} (\bibinfo {year} {2001})}\BibitemShut {NoStop}%
		\bibitem [{\citenamefont {Roberts}\ \emph {et~al.}(2002)\citenamefont
			{Roberts}, \citenamefont {Schmidt},\ and\ \citenamefont
			{Vinnik}}]{roberts2002quantum}%
		\BibitemOpen
		\bibfield  {author} {\bibinfo {author} {\bibfnamefont {C.}~\bibnamefont
				{Roberts}}, \bibinfo {author} {\bibfnamefont {S.}~\bibnamefont {Schmidt}},\
			and\ \bibinfo {author} {\bibfnamefont {D.}~\bibnamefont {Vinnik}},\
		}\bibfield  {title} {\bibinfo {title} {Quantum effects with an x-ray
				free-electron laser},\ }\href {https://doi.org/10.1103/PhysRevLett.89.153901}
		{\bibfield  {journal} {\bibinfo  {journal} {Phys. Rev. Lett.}\ }\textbf
			{\bibinfo {volume} {89}},\ \bibinfo {pages} {153901} (\bibinfo {year}
			{2002})}\BibitemShut {NoStop}%
		\bibitem [{\citenamefont {Prakapenia}\ and\ \citenamefont
			{Vereshchagin}(2023)}]{prakapenia2023pauli}%
		\BibitemOpen
		\bibfield  {author} {\bibinfo {author} {\bibfnamefont {M.}~\bibnamefont
				{Prakapenia}}\ and\ \bibinfo {author} {\bibfnamefont {G.}~\bibnamefont
				{Vereshchagin}},\ }\bibfield  {title} {\bibinfo {title} {Pauli blocking
				effects on pair creation in strong electric field},\ }\href
		{https://doi.org/10.1103/PhysRevD.108.013002} {\bibfield  {journal} {\bibinfo
				{journal} {Phys. Rev. D}\ }\textbf {\bibinfo {volume} {108}},\ \bibinfo
			{pages} {013002} (\bibinfo {year} {2023})}\BibitemShut {NoStop}%
		\bibitem [{\citenamefont {Jiang}\ \emph {et~al.}(2023)\citenamefont {Jiang},
			\citenamefont {Gong}, \citenamefont {Li},\ and\ \citenamefont
			{Li}}]{jiang2023backreaction}%
		\BibitemOpen
		\bibfield  {author} {\bibinfo {author} {\bibfnamefont {R.~Z.}\ \bibnamefont
				{Jiang}}, \bibinfo {author} {\bibfnamefont {C.}~\bibnamefont {Gong}},
			\bibinfo {author} {\bibfnamefont {Z.~L.}\ \bibnamefont {Li}},\ and\ \bibinfo
			{author} {\bibfnamefont {Y.~J.}\ \bibnamefont {Li}},\ }\bibfield  {title}
		{\bibinfo {title} {Backreaction effect and plasma oscillation in pair
				production for rapidly oscillating electric fields},\ }\href
		{https://doi.org/10.1103/PhysRevD.108.076015} {\bibfield  {journal} {\bibinfo
				{journal} {Phys. Rev. D}\ }\textbf {\bibinfo {volume} {108}},\ \bibinfo
			{pages} {076015} (\bibinfo {year} {2023})}\BibitemShut {NoStop}%
		\bibitem [{\citenamefont {Keldysh}(1965)}]{keldysh1965ionization}%
		\BibitemOpen
		\bibfield  {author} {\bibinfo {author} {\bibfnamefont {L.~V.}\ \bibnamefont
				{Keldysh}},\ }\bibfield  {title} {\bibinfo {title} {Ionization in the field
				of a strong electromagnetic wave},\ }\href
		{https://doi.org/10.1142/9789811279461_0008} {\bibfield  {journal} {\bibinfo
				{journal} {Sov. Phys. JETP}\ }\textbf {\bibinfo {volume} {20}},\ \bibinfo
			{pages} {1307} (\bibinfo {year} {1965})}\BibitemShut {NoStop}%
		\bibitem [{\citenamefont {Br{\'e}zin}\ and\ \citenamefont
			{Itzykson}(1970)}]{brezin1970pair}%
		\BibitemOpen
		\bibfield  {author} {\bibinfo {author} {\bibfnamefont {E.}~\bibnamefont
				{Br{\'e}zin}}\ and\ \bibinfo {author} {\bibfnamefont {C.}~\bibnamefont
				{Itzykson}},\ }\bibfield  {title} {\bibinfo {title} {Pair production in
				vacuum by an alternating field},\ }\href
		{https://doi.org/10.1103/PhysRevD.2.1191} {\bibfield  {journal} {\bibinfo
				{journal} {Phys. Rev. D}\ }\textbf {\bibinfo {volume} {2}},\ \bibinfo {pages}
			{1191} (\bibinfo {year} {1970})}\BibitemShut {NoStop}%
		\bibitem [{\citenamefont {Popov}(2004)}]{popov2004Keldysh}%
		\BibitemOpen
		\bibfield  {author} {\bibinfo {author} {\bibfnamefont {V.~S.}\ \bibnamefont
				{Popov}},\ }\bibfield  {title} {\bibinfo {title} {Tunnel and multiphoton
				ionization of atoms and ions in a strong laser field (keldysh theory)},\
		}\href {https://ufn.ru/en/articles/2004/9/a/} {\bibfield  {journal} {\bibinfo
				{journal} {Phys. Usp.}\ }\textbf {\bibinfo {volume} {47}},\ \bibinfo {pages}
			{855} (\bibinfo {year} {2004})}\BibitemShut {NoStop}%
		\bibitem [{\citenamefont {Aleksandrov}\ \emph
			{et~al.}(2017{\natexlab{a}})\citenamefont {Aleksandrov}, \citenamefont
			{Plunien},\ and\ \citenamefont {Shabaev}}]{aleksandrov2017dipole}%
		\BibitemOpen
		\bibfield  {author} {\bibinfo {author} {\bibfnamefont {I.~A.}\ \bibnamefont
				{Aleksandrov}}, \bibinfo {author} {\bibfnamefont {G.}~\bibnamefont
				{Plunien}},\ and\ \bibinfo {author} {\bibfnamefont {V.~M.}\ \bibnamefont
				{Shabaev}},\ }\bibfield  {title} {\bibinfo {title} {Momentum distribution of
				particles created in space-time-dependent colliding laser pulses},\ }\href
		{https://doi.org/10.1103/PhysRevD.96.076006} {\bibfield  {journal} {\bibinfo
				{journal} {Phys. Rev. D}\ }\textbf {\bibinfo {volume} {96}},\ \bibinfo
			{pages} {076006} (\bibinfo {year} {2017}{\natexlab{a}})}\BibitemShut
		{NoStop}%
		\bibitem [{\citenamefont {Blinne}\ and\ \citenamefont
			{Gies}(2014)}]{blinne2014rotating}%
		\BibitemOpen
		\bibfield  {author} {\bibinfo {author} {\bibfnamefont {A.}~\bibnamefont
				{Blinne}}\ and\ \bibinfo {author} {\bibfnamefont {H.}~\bibnamefont {Gies}},\
		}\bibfield  {title} {\bibinfo {title} {Pair production in rotating electric
				fields},\ }\href {https://doi.org/10.1103/PhysRevD.89.085001} {\bibfield
			{journal} {\bibinfo  {journal} {Phys. Rev. D}\ }\textbf {\bibinfo {volume}
				{89}},\ \bibinfo {pages} {085001} (\bibinfo {year} {2014})}\BibitemShut
		{NoStop}%
		\bibitem [{\citenamefont {Aleksandrov}\ and\ \citenamefont
			{Kudlis}(2024)}]{aleksandrov2024rotating}%
		\BibitemOpen
		\bibfield  {author} {\bibinfo {author} {\bibfnamefont {I.~A.}\ \bibnamefont
				{Aleksandrov}}\ and\ \bibinfo {author} {\bibfnamefont {A.}~\bibnamefont
				{Kudlis}},\ }\bibfield  {title} {\bibinfo {title} {Pair production in
				rotating electric fields via quantum kinetic equations: Resolving helicity
				states},\ }\href {https://doi.org/10.1103/PhysRevD.110.L011901} {\bibfield
			{journal} {\bibinfo  {journal} {Phys. Rev. D}\ }\textbf {\bibinfo {volume}
				{110}},\ \bibinfo {pages} {L011901} (\bibinfo {year} {2024})}\BibitemShut
		{NoStop}%
		\bibitem [{\citenamefont {Adorno}\ \emph {et~al.}(2015)\citenamefont {Adorno},
			\citenamefont {Gavrilov},\ and\ \citenamefont
			{Gitman}}]{adorno2015switchoff}%
		\BibitemOpen
		\bibfield  {author} {\bibinfo {author} {\bibfnamefont {T.~C.}\ \bibnamefont
				{Adorno}}, \bibinfo {author} {\bibfnamefont {S.~P.}\ \bibnamefont
				{Gavrilov}},\ and\ \bibinfo {author} {\bibfnamefont {D.~M.}\ \bibnamefont
				{Gitman}},\ }\bibfield  {title} {\bibinfo {title} {Particle creation from the
				vacuum by an exponentially decreasing electric field},\ }\href
		{https://doi.org/10.1088/0031-8949/90/7/074005} {\bibfield  {journal}
			{\bibinfo  {journal} {Physica Scripta}\ }\textbf {\bibinfo {volume} {90}},\
			\bibinfo {pages} {074005} (\bibinfo {year} {2015})}\BibitemShut {NoStop}%
		\bibitem [{\citenamefont {Aleksandrov}\ \emph {et~al.}(2018)\citenamefont
			{Aleksandrov}, \citenamefont {Plunien},\ and\ \citenamefont
			{Shabaev}}]{aleksandrov2018dase}%
		\BibitemOpen
		\bibfield  {author} {\bibinfo {author} {\bibfnamefont {I.~A.}\ \bibnamefont
				{Aleksandrov}}, \bibinfo {author} {\bibfnamefont {G.}~\bibnamefont
				{Plunien}},\ and\ \bibinfo {author} {\bibfnamefont {V.~M.}\ \bibnamefont
				{Shabaev}},\ }\bibfield  {title} {\bibinfo {title} {Dynamically assisted
				schwinger effect beyond the spatially-uniform-field approximation},\ }\href
		{https://doi.org/10.1103/PhysRevD.97.116001} {\bibfield  {journal} {\bibinfo
				{journal} {Phys. Rev. D}\ }\textbf {\bibinfo {volume} {97}},\ \bibinfo
			{pages} {116001} (\bibinfo {year} {2018})}\BibitemShut {NoStop}%
		\bibitem [{\citenamefont {Aleksandrov}\ and\ \citenamefont
			{Kohlf\"urst}(2020)}]{aleksandrov2020spacetime}%
		\BibitemOpen
		\bibfield  {author} {\bibinfo {author} {\bibfnamefont {I.~A.}\ \bibnamefont
				{Aleksandrov}}\ and\ \bibinfo {author} {\bibfnamefont {C.}~\bibnamefont
				{Kohlf\"urst}},\ }\bibfield  {title} {\bibinfo {title} {Pair production in
				temporally and spatially oscillating fields},\ }\href
		{https://doi.org/10.1103/PhysRevD.101.096009} {\bibfield  {journal} {\bibinfo
				{journal} {Phys. Rev. D}\ }\textbf {\bibinfo {volume} {101}},\ \bibinfo
			{pages} {096009} (\bibinfo {year} {2020})}\BibitemShut {NoStop}%
		\bibitem [{\citenamefont {Kohlf\"urst}(2024)}]{kohlfurst2024}%
		\BibitemOpen
		\bibfield  {author} {\bibinfo {author} {\bibfnamefont {C.}~\bibnamefont
				{Kohlf\"urst}},\ }\bibfield  {title} {\bibinfo {title} {Pair production in
				circularly polarized waves},\ }\href
		{https://doi.org/10.1103/PhysRevD.110.L111903} {\bibfield  {journal}
			{\bibinfo  {journal} {Phys. Rev. D}\ }\textbf {\bibinfo {volume} {110}},\
			\bibinfo {pages} {L111903} (\bibinfo {year} {2024})}\BibitemShut {NoStop}%
		\bibitem [{\citenamefont {Aleksandrov}\ \emph
			{et~al.}(2017{\natexlab{b}})\citenamefont {Aleksandrov}, \citenamefont
			{Plunien},\ and\ \citenamefont {Shabaev}}]{aleksandrov2017shape}%
		\BibitemOpen
		\bibfield  {author} {\bibinfo {author} {\bibfnamefont {I.~A.}\ \bibnamefont
				{Aleksandrov}}, \bibinfo {author} {\bibfnamefont {G.}~\bibnamefont
				{Plunien}},\ and\ \bibinfo {author} {\bibfnamefont {V.~M.}\ \bibnamefont
				{Shabaev}},\ }\bibfield  {title} {\bibinfo {title} {Pulse shape effects on
				the electron-positron pair production in strong laser fields},\ }\href
		{https://doi.org/10.1103/PhysRevD.95.056013} {\bibfield  {journal} {\bibinfo
				{journal} {Phys. Rev. D}\ }\textbf {\bibinfo {volume} {95}},\ \bibinfo
			{pages} {056013} (\bibinfo {year} {2017}{\natexlab{b}})}\BibitemShut
		{NoStop}%
		\bibitem [{\citenamefont {Abdukerim}\ \emph {et~al.}(2013)\citenamefont
			{Abdukerim}, \citenamefont {Li},\ and\ \citenamefont
			{Xie}}]{abdukerim2013effects}%
		\BibitemOpen
		\bibfield  {author} {\bibinfo {author} {\bibfnamefont {N.}~\bibnamefont
				{Abdukerim}}, \bibinfo {author} {\bibfnamefont {Z.-L.}\ \bibnamefont {Li}},\
			and\ \bibinfo {author} {\bibfnamefont {B.-S.}\ \bibnamefont {Xie}},\
		}\bibfield  {title} {\bibinfo {title} {Effects of laser pulse shape and
				carrier envelope phase on pair production},\ }\href
		{https://doi.org/10.1016/j.physletb.2013.09.014} {\bibfield  {journal}
			{\bibinfo  {journal} {Phys. Lett. B}\ }\textbf {\bibinfo {volume} {726}},\
			\bibinfo {pages} {820} (\bibinfo {year} {2013})}\BibitemShut {NoStop}%
		\bibitem [{\citenamefont {Otto}\ \emph {et~al.}(2018)\citenamefont {Otto},
			\citenamefont {Oppitz},\ and\ \citenamefont
			{K{\"a}mpfer}}]{otto2018assisted}%
		\BibitemOpen
		\bibfield  {author} {\bibinfo {author} {\bibfnamefont {A.}~\bibnamefont
				{Otto}}, \bibinfo {author} {\bibfnamefont {H.}~\bibnamefont {Oppitz}},\ and\
			\bibinfo {author} {\bibfnamefont {B.}~\bibnamefont {K{\"a}mpfer}},\
		}\bibfield  {title} {\bibinfo {title} {Assisted vacuum decay by
				time-dependent electric fields},\ }\href
		{https://doi.org/10.1140/epja/i2018-12473-x} {\bibfield  {journal} {\bibinfo
				{journal} {Eur. Phys. J. A}\ }\textbf {\bibinfo {volume} {54}},\ \bibinfo
			{pages} {23} (\bibinfo {year} {2018})}\BibitemShut {NoStop}%
		\bibitem [{\citenamefont {Panferov}\ \emph {et~al.}(2016)\citenamefont
			{Panferov} \emph {et~al.}}]{panferov2016assisted}%
		\BibitemOpen
		\bibfield  {author} {\bibinfo {author} {\bibfnamefont {A.~D.}\ \bibnamefont
				{Panferov}} \emph {et~al.},\ }\bibfield  {title} {\bibinfo {title} {Assisted
				dynamical schwinger effect: pair production in a pulsed bifrequent field},\
		}\href {https://doi.org/10.1140/epjd/e2016-60517-y} {\bibfield  {journal}
			{\bibinfo  {journal} {Eur. Phys. J. D}\ }\textbf {\bibinfo {volume} {70}},\
			\bibinfo {pages} {56} (\bibinfo {year} {2016})}\BibitemShut {NoStop}%
		\bibitem [{\citenamefont {Bra{\ss}}\ \emph {et~al.}(2025)\citenamefont
			{Bra{\ss}} \emph {et~al.}}]{BraS2025}%
		\BibitemOpen
		\bibfield  {author} {\bibinfo {author} {\bibfnamefont {J.}~\bibnamefont
				{Bra{\ss}}} \emph {et~al.},\ }\href@noop {} {\bibinfo {title} {Relative-phase
				dependence of dynamically assisted electron-positron pair creation in the
				superposition of strong oscillating electric-field pulses}} (\bibinfo {year}
		{2025}),\ \Eprint {https://arxiv.org/abs/2505.24488} {arXiv:2505.24488
			[hep-ph]} \BibitemShut {NoStop}%
		\bibitem [{\citenamefont {Hebenstreit}\ and\ \citenamefont
			{Fillion-Gourdeau}(2014)}]{hebestreit2014optimal}%
		\BibitemOpen
		\bibfield  {author} {\bibinfo {author} {\bibfnamefont {F.}~\bibnamefont
				{Hebenstreit}}\ and\ \bibinfo {author} {\bibfnamefont {F.}~\bibnamefont
				{Fillion-Gourdeau}},\ }\bibfield  {title} {\bibinfo {title} {Optimization of
				schwinger pair production in colliding laser pulses},\ }\href
		{https://doi.org/10.1016/j.physletb.2014.10.056} {\bibfield  {journal}
			{\bibinfo  {journal} {Phys. Lett. B.}\ }\textbf {\bibinfo {volume} {739}},\
			\bibinfo {pages} {189} (\bibinfo {year} {2014})}\BibitemShut {NoStop}%
		\bibitem [{\citenamefont {Mohamedsedik}\ \emph {et~al.}(2023)\citenamefont
			{Mohamedsedik}, \citenamefont {Li}, \citenamefont {Wang}, \citenamefont
			{Amat}, \citenamefont {Hu},\ and\ \citenamefont
			{Xie}}]{mohamedsedik2023phase}%
		\BibitemOpen
		\bibfield  {author} {\bibinfo {author} {\bibfnamefont {M.}~\bibnamefont
				{Mohamedsedik}}, \bibinfo {author} {\bibfnamefont {L.-J.}\ \bibnamefont
				{Li}}, \bibinfo {author} {\bibfnamefont {L.}~\bibnamefont {Wang}}, \bibinfo
			{author} {\bibfnamefont {O.}~\bibnamefont {Amat}}, \bibinfo {author}
			{\bibfnamefont {L.-N.}\ \bibnamefont {Hu}},\ and\ \bibinfo {author}
			{\bibfnamefont {B.}~\bibnamefont {Xie}},\ }\bibfield  {title} {\bibinfo
			{title} {Phase effect on and symmetry of pair production in inhomogeneous
				electric fields with chirping},\ }\href
		{https://doi.org/10.1140/epjp/s13360-023-03926-1} {\bibfield  {journal}
			{\bibinfo  {journal} {Eur. Phys. J. Plus}\ }\textbf {\bibinfo {volume}
				{138}},\ \bibinfo {pages} {316} (\bibinfo {year} {2023})}\BibitemShut
		{NoStop}%
		\bibitem [{\citenamefont {Mohamedsedik}\ \emph {et~al.}(2021)\citenamefont
			{Mohamedsedik}, \citenamefont {Li},\ and\ \citenamefont
			{Xie}}]{mohamedsedik2021schwinger}%
		\BibitemOpen
		\bibfield  {author} {\bibinfo {author} {\bibfnamefont {M.}~\bibnamefont
				{Mohamedsedik}}, \bibinfo {author} {\bibfnamefont {L.-J.}\ \bibnamefont
				{Li}},\ and\ \bibinfo {author} {\bibfnamefont {B.}~\bibnamefont {Xie}},\
		}\bibfield  {title} {\bibinfo {title} {Schwinger pair production in
				inhomogeneous electric fields with symmetrical frequency chirp},\ }\href
		{https://doi.org/10.1103/PhysRevD.104.016009} {\bibfield  {journal} {\bibinfo
				{journal} {Phys. Rev. D}\ }\textbf {\bibinfo {volume} {104}},\ \bibinfo
			{pages} {016009} (\bibinfo {year} {2021})}\BibitemShut {NoStop}%
		\bibitem [{\citenamefont {Dumlu}\ and\ \citenamefont
			{Dunne}(2011{\natexlab{a}})}]{dumlu2011interference}%
		\BibitemOpen
		\bibfield  {author} {\bibinfo {author} {\bibfnamefont {C.~K.}\ \bibnamefont
				{Dumlu}}\ and\ \bibinfo {author} {\bibfnamefont {G.~V.}\ \bibnamefont
				{Dunne}},\ }\bibfield  {title} {\bibinfo {title} {Interference effects in
				schwinger vacuum pair production for time-dependent laser pulses},\ }\href
		{https://doi.org/10.1103/PhysRevD.83.065028} {\bibfield  {journal} {\bibinfo
				{journal} {Phys. Rev. D}\ }\textbf {\bibinfo {volume} {83}},\ \bibinfo
			{pages} {065028} (\bibinfo {year} {2011}{\natexlab{a}})}\BibitemShut
		{NoStop}%
		\bibitem [{\citenamefont {Akkermans}\ and\ \citenamefont
			{Dunne}(2012)}]{akkermans2012ramsey}%
		\BibitemOpen
		\bibfield  {author} {\bibinfo {author} {\bibfnamefont {E.}~\bibnamefont
				{Akkermans}}\ and\ \bibinfo {author} {\bibfnamefont {G.~V.}\ \bibnamefont
				{Dunne}},\ }\bibfield  {title} {\bibinfo {title} {Ramsey fringes and
				time-domain multiple-slit interference from vacuum},\ }\href
		{https://doi.org/10.1103/PhysRevLett.108.030401} {\bibfield  {journal}
			{\bibinfo  {journal} {Phys. Rev. Lett.}\ }\textbf {\bibinfo {volume} {108}},\
			\bibinfo {pages} {030401} (\bibinfo {year} {2012})}\BibitemShut {NoStop}%
		\bibitem [{\citenamefont {Taya}\ \emph {et~al.}(2021)\citenamefont {Taya},
			\citenamefont {Fujimori}, \citenamefont {Misumi}, \citenamefont {Nitta},\
			and\ \citenamefont {Sakai}}]{taya2021exact}%
		\BibitemOpen
		\bibfield  {author} {\bibinfo {author} {\bibfnamefont {H.}~\bibnamefont
				{Taya}}, \bibinfo {author} {\bibfnamefont {T.}~\bibnamefont {Fujimori}},
			\bibinfo {author} {\bibfnamefont {T.}~\bibnamefont {Misumi}}, \bibinfo
			{author} {\bibfnamefont {M.}~\bibnamefont {Nitta}},\ and\ \bibinfo {author}
			{\bibfnamefont {N.}~\bibnamefont {Sakai}},\ }\bibfield  {title} {\bibinfo
			{title} {Exact wkb analysis of the vacuum pair production by time-dependent
				electric fields},\ }\href {https://doi.org/10.1007/JHEP03(2021)082}
		{\bibfield  {journal} {\bibinfo  {journal} {J. High Energ. Phys.}\ }\textbf
			{\bibinfo {volume} {2021}}\bibinfo  {number} { (3)},\ \bibinfo {pages}
			{1}}\BibitemShut {NoStop}%
		\bibitem [{\citenamefont {Dumlu}\ and\ \citenamefont
			{Dunne}(2011{\natexlab{b}})}]{dumlu2011complex}%
		\BibitemOpen
		\bibfield  {number} {  }\bibfield  {author} {\bibinfo {author} {\bibfnamefont
				{C.~K.}\ \bibnamefont {Dumlu}}\ and\ \bibinfo {author} {\bibfnamefont
				{G.~V.}\ \bibnamefont {Dunne}},\ }\bibfield  {title} {\bibinfo {title}
			{Complex worldline instantons and quantum interference in vacuum pair
				production},\ }\href {https://doi.org/10.1103/PhysRevD.84.125023} {\bibfield
			{journal} {\bibinfo  {journal} {Phys. Rev. D}\ }\textbf {\bibinfo {volume}
				{84}},\ \bibinfo {pages} {125023} (\bibinfo {year}
			{2011}{\natexlab{b}})}\BibitemShut {NoStop}%
	\end{thebibliography}
	
	\providecommand{\noopsort}[1]{}\providecommand{\singleletter}[1]{#1}%

\end{document}